\documentclass[journal, onecolumn]{IEEEtran}
\IEEEoverridecommandlockouts

\usepackage{amsmath, mathrsfs, graphicx,epsfig,amssymb, amsfonts, color, subfigure}
\usepackage[all]{xy}
\usepackage[center]{caption}

\newtheorem{lemma}{\bf Lemma}

\newtheorem{defn}{\bf Definition}
\newtheorem{thm}{\bf Theorem}

\newtheorem{proof}{\bf Proof}

\newtheorem{rem}{\bf Remark}
\newtheorem{cor}{\bf Corollary}
\newtheorem{ex}{\bf Example}

\begin{document}

\title{The Capacity Region of the MIMO Interference Channel and its Reciprocity to Within a Constant Gap }

\author{{\Large Sanjay ~Karmakar ~~~~~~Mahesh~ K. ~Varanasi}
\thanks{S. Karmakar was with the Department
of Electrical, Computer and Energy Engineering, University of Colorado, Boulder,
CO, 80309 USA. He is now with the Department
of Electrical and Computer Engineering, North Dakota State University, Fargo,
ND, 58108, USA. e-mail: sanjay.karmakar@ndsu.edu.}
\thanks{ Mahesh K. Varanasi is with the Department
of Electrical, Computer and Energy Engineering, University of Colorado, Boulder,
CO, 80309 USA. e-mail: varanasi@colorado.edu.}
\thanks{The material in this paper was presented in part at the 2010 IEEE International Symposium on Information Theory, Austin, TX and the 2011 IEEE International Symposium on Information Theory, St. Petersburg, Russia.}
}

\maketitle


\begin{abstract}
The capacity region of the $2$-user multi-input multi-output (MIMO) Gaussian interference channel (IC) is characterized to within a constant gap that is independent of the 
channel matrices for the general case of the MIMO IC with an arbitrary number of antennas at each node. An achievable rate region and an outer bound to the capacity region of a class of interference channels were obtained in previous work by Telatar and Tse as unions over all possible input distributions. In contrast to that previous work on the MIMO IC, a simple and an explicit achievable coding scheme are obtained here and shown to have the constant-gap-to-capacity property and in which the sub-rates of the common and private messages of each user are explicitly specified for each achievable rate pair. The constant-gap-to-capacity results are thus proved in this work by first establishing explicit upper and lower bounds to the capacity region. A reciprocity result is also proved which is that the capacity of the reciprocal MIMO IC is within a constant gap of the capacity region of the forward MIMO IC.
\end{abstract}

\begin{IEEEkeywords}
Capacity, Interference channel, Multi-input, multi-output (MIMO), Reciprocity.
\end{IEEEkeywords}


%
\IEEEpeerreviewmaketitle


\section{Introduction}
The two-user interference channel is a model for a single-hop, multi-flow wireless network in which multiple transmit-receive pairs communicate over a common noisy channel in that it captures the fundamental interactions between the multiple transmitted signals in such networks, namely, {\it broadcast}, {\it superposition} and {\it interference}. This model was first mentioned in \cite{Shannon}, and was studied in a series of works in \cite{Carleial75,Sato,Han_Kobayashi,Costa_Gamal,Sato78,Benzel,Gamal_Costa} that considered certain special classes of the IC where the capacity regions of the so-called {\em very strong} IC, the {\em strong} IC and certain classes of {\em degraded} and {\em deterministic} ICs, respectively, were established. Different sets of inner and outer bounds considering the embedded multiple-access and broadcast and Z channels were derived in \cite{Sato77,Sato78,Carleial78,Carleial83,Costa85,Kramer2004}. However, the Han-Kobayashi (HK) \cite{Han_Kobayashi} coding scheme that improves that of the rate-splitting strategy of \cite{Carleial83} remains the best known achievable scheme for this channel. In spite of over 3 decades of research, the capacity region in the general case remained unsolved.

Recent results include the simplified description of the HK rate region due to Chong et al. in~\cite{CMG} (see also \cite{KH2007,Hodtani}) and the capacity regions of new and/or more general classes of channels than for which capacity was previously known, e.g., the sum capacity of the so-called {\it noisy} interference channels was found in \cite{Annapureddy_Veeravalli,Motahari_Khandani,Shang2009} and the capacity region of the {\it very strong} and {\it aligned strong} MIMO IC were found in \cite{SCKP}. The common feature of this line of work is that it focuses on a small subset of channel parameters but seeks to solve the challenging problem of obtaining the exact capacity of the channel.

A different line of research was initiated by Etkin et al.~\cite{ETW1}, where the authors find an approximation of the capacity region of the two-user scalar Gaussian IC where the criterion of approximation is to specify the capacity region to within a constant gap independently of SNR and the direct and cross channel coefficients. Moreover, they obtain that result through a {\em simple} HK scheme, i.e., by identifying a single, channel parameter dependent, joint distribution of input and auxiliary random variables among the infinitely many possible specifications including time-sharing that together contribute to the general HK rate region. The key feature of this simple HK scheme is that each user employs independent Gaussian superposition coding of private and public messages with the private message power set so that it reaches the unintended receiver at the noise level. A one bit gap to capacity was proved in \cite{ETW1} using the simplified description of the HK rate region of \cite{CMG}. Thus, the result of \cite{ETW1} characterizes the capacity region to within a constant gap that is independent of the SNR and all channel coefficients. Moreover, it identifies a simple HK scheme that has this property, thereby also providing an explicit expression for the achievable rate region in terms of channel parameters.

Since most modern wireless communication systems feature multiple antennas at some or all terminals it is of interest to study the two-user Gaussian MIMO IC. However, multiple antennas at different nodes make it harder to obtain results similar to those available for the SISO IC. For instance, the deterministic model developed in \cite{Bresler_Tse} for the two-user SISO IC which was shown to reproduce the constant gap to capacity approximation result of \cite{ETW1} doesn't extend to MIMO channels. Moreover, as compared to the result on the capacity of the strong SISO IC, the capacity of the MIMO IC is known~\cite{SCKP} only for the so-called {\it aligned strong} interference regime, where the direct and cross link channel matrices satisfy a matrix equation. Such a matrix equation between the two channel matrices may seldom, if ever, hold. In general, the problem of characterizing the exact capacity of a MIMO IC even for small and special classes can be challenging; this point is also illustrated by \cite{Sriram_Jafar} where the capacity region of a class of {\it very strong} MISO ICs was characterized.

In \cite{TT}, Telatar and Tse consider an interesting class of two-user semi-deterministic discrete memoryless ICs which generalizes the class of deterministic ICs of \cite{Gamal_Costa} and is also applicable to the Gaussian MIMO IC. They obtain outer bounds to the capacity region that are within a gap specified in terms of certain conditional mutual informations to the general HK achievable region~\cite{CMG}. The implication of this work to the two-user MIMO IC is that the union of all the achievable rate sub-regions of the general HK scheme (one sub-region for each input distribution), is within a constant gap (of $N_i$ bits, where $N_i$ is the number of antennas at receiver $i$) to the outer bound developed therein (which in turn is given as a union over all input and time-sharing distributions), and hence, to the capacity region. However, no specific achievable scheme is identified with the constant-gap-to-capacity property among the infinitely many possibilities that make up the the general HK scheme. In fact, it is unclear from that work if there exists a {\em simple} HK scheme in general (corresponding to a single input distribution, as it does for the SISO case \cite{ETW1} ) or even an {\em explicit} HK scheme (whose rate region is the union of rate regions achievable by a finite number of input distributions) with the constant-gap-to-capacity property. Moreover, since the upper and lower bounds are not given explicitly as functions of the channel matrices in \cite{TT} they cannot be used for further analysis such as, for example, for finding the generalized degrees of freedom (GDoF) analysis, as mentioned in \cite{Wang_Tse}.

In this paper, we consider the two-user Gaussian MIMO IC with an arbitrary number of antennas at each node. Without restricting the channel matrices in
any way, we obtain constant-gap-to-capacity characterizations through a simple and an explicit HK scheme, neither of which involves time-sharing.
The approach we adopt is as follows: first, we establish a set of {\em explicit} channel-matrices-dependent upper bounds to the capacity region of the $2$-user MIMO IC under input power constraints, i.e., the resulting explicit outer bound on the capacity region does not involve a union over input distributions as does the outer bound in \cite{TT}. Consequently, inspired by a novel interpretation of this outer bound, we propose a simple HK coding scheme which involves independent Gaussian linear superposition coding with certain explicit channel dependent covariance matrix assignments for the private and public messages of each user and show that this input distribution produces a rate region that is within a constant gap to the capacity region. Moreover, the explicit bounds obtained here were used by the authors to obtain the GDoF region of the MIMO IC in the companion paper \cite{Karmakar-Varanasi-GDoF}.

The above specification of coding scheme does not conform to the specification of \cite{TT} on the choice of the conditional distribution of the auxiliary random variables given the inputs. It is thus distinct from any achievable scheme that might result as a consequence of the prescription of \cite{TT} (see the discussion following Theorem \ref{thm_approximate_capacity}). Moreover, since in the HK coding scheme the public message of a user gets decoded at the receiver of the other user, it is important to choose the sub-rates of the private and public messages of each user carefully because an arbitrary rate for the public message might not be supported if the corresponding cross-link is weak. We thus also specify explicitly the set of these sub-rates for the private and public messages for which such a scenario never arises. In fact, a two-dimensional projection of this latter set actually yields the achievable rate region of the simple HK coding scheme.

The gap to capacity of the aforementioned simple HK coding scheme is then improved by proposing an explicit HK scheme where the transmitters are allowed to use one of three simple superposition coding schemes depending on the operating rate pair. This explicit scheme is inspired by the recent result in \cite{CMG} that proves that such a union contains the so-called compact rate region that is larger than the HK achievable rate region for a particular choice of distributions of the inputs, auxiliary random variables and time sharing random variable.
Interestingly, for a large class of MIMO ICs, this latter gap is the same as that
reported in \cite{TT}\footnote{Although it was claimed in \cite{TT} that the gap to capacity obtained therein is within $(N_1, N_2)$ bits it can be shown that the gap is bounded by $(\min\{N_1,M_2\}, \min\{N_2,M_1\})$ bits.}. 
This class includes, for example, single-input, multiple-output (SIMO) ICs with single-antenna transmitters and multiple antenna receivers) for which the gap is one bit.

Using the explicit expressions for both the achievable rate region and the set of upper bounds to the capacity region of the MIMO IC, we then derive an interesting {\it reciprocity} result which is that the capacity of a two-user MIMO IC is within a constant gap to that of the channel obtained by interchanging the roles of the transmitters and the receivers. 

The rest of the paper is organized as follows. Following a description of the notations used in this paper we specify the system model in Section~\ref{sec_system_model_and_preliminaries}. In Section~\ref{sec_approximate_capacity}, we derive a set of upper bounds to the capacity region and two different rate regions achievable by one simple and one explicit HK coding scheme. Comparing the set of upper and lower bounds, the capacity region of the MIMO IC is characterized within a constant number of bits. As a byproduct of this analysis, we also prove the {\it reciprocity} of the capacity region of the MIMO IC in the approximate capacity sense in Section~\ref{subsection_reciprocity}. Finally, Section~\ref{sec_conclusion} concludes the paper. In order that the paper is easy to read, many of the proofs are given in Appendices.

\begin{proof}[Notations]
Let $\mathbb{C}$ and $\mathbb{R}^+$ represent the field of complex numbers and the set of non-negative real numbers, respectively. An $n\times m$ matrix with entries in $\mathbb{C}$ will be denoted as $A\in \mathbb{C}^{n\times m}$. The conjugate transpose of the matrix $A$ is denoted as $A^{\dagger}$ and its determinant as $|A|$. Let $\|z\|^2$ represents the square of the absolute value of the complex number, i.e., if $z=(x+iy)$ then $\|z\|^2=x^2+y^2$. The trace of the matrix $A\in \mathbb{C}^{n\times n}$ is denoted as $\textrm{Tr}(A)$, i.e., $\textrm{Tr}(A)=\sum_{i=1}^{n}a_{ii}$, where $a_{ij}$ represents the element in the $i$-th row and $j$-th column of the matrix $A$. The Frobenius norm of the matrix $A$ is denote by $\|A\|_F^2$, i.e., $\|A\|_F^2=\textrm{Tr}(AA^\dagger)$. $I_n$ represents the $n\times n$ identity matrix, $0_{m\times n}$ represents an all zero $m\times n$ matrix and $\mathbb{U}^{n\times n}$ represents the set of $n\times n$ unitary matrices. The $k^{th}$ column of the matrix $A$ will be denoted by $A^{[k]}$ whereas $A^{[k_1:k_2]}$ represents a matrix whose columns are same as the $k_1^{th}$ to $k_2^{th}$ columns of matrix $A$. $ | \mathcal{A}| $ denotes the cardinality of the set $ \mathcal{A} $. The fact that $(A-B)$ is a positive semi-definite (p.s.d.) (or positive definite (p.d.)) matrix is denoted by $A\succeq B$ (or $A\succ B$). $A\otimes B$ denotes the tensor or Kronecker product of the two matrices. If $x_{t}\in \mathbb{C}^{m\times 1}, \forall \; 1\leq t\leq n$, then $x^n\triangleq [x_1^{\dagger},\cdots , x_n^{\dagger}]^{\dagger}$. $\{A,B,C,D\}$ will represent an ordered set of matrices. Moreover, $I(X;Y), I(X;Y|Z), h(X)$ and $h(X|Y)$ represents the mutual information, conditional mutual information, differential entropy and conditional differential entropy of the random variable arguments, respectively. The quantities $x\land y$, $x\vee y$ and $(x)^+$ (or $[x]^+$) denote the minimum and maximum between $x$ and $y$ and the $\max\{x,0\}$, respectively. All the logarithms in this paper are with base $2$. The distribution of a complex circularly symmetric Gaussian random vector with zero mean and covariance matrix $Q$ is denoted as $\mathcal{CN}(0,Q)$.
\end{proof}

\section{Channel Model and Mathematical preliminaries }
\label{sec_system_model_and_preliminaries}

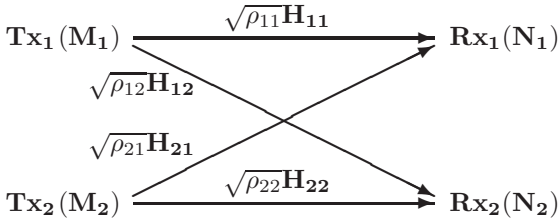
\begin{figure}[!thb]
\setlength{\unitlength}{1mm}
\begin{picture}(80,40)
\thicklines
\put(20,10){\vector(1,0){40}}
\put(20,32){\vector(1,0){40}}
\put(20,11){\vector(2,1){40}}
\put(20,31){\vector(2,-1){40}}
\put(10,9){$\mathbf{(M_2)}$}
\put(69,9){$\mathbf{(N_2)}$}
\put(10,31){$\mathbf{(M_1)}$}
\put(69,31){$\mathbf{(N_1)}$}

\put(3,9){$\mathbf{Tx_2}$}
\put(62,9){$\mathbf{Rx_2}$}
\put(3,31){$\mathbf{Tx_1}$}
\put(62,31){$\mathbf{Rx_1}$}

\put(32,12){$\sqrt{\rho_{22}}\mathbf{H_{22}}$}
\put(32,34){$\sqrt{\rho_{11}}\mathbf{H_{11}}$}
\put(14,17){$\sqrt{\rho_{21}}\mathbf{H_{21}}$}
\put(14,25){$\sqrt{\rho_{12}}\mathbf{H_{12}}$}
\end{picture}
\caption{The $(M_1,N_1,M_2,N_2)$ MIMO IC.}
\label{channel_model_two_user_IC}
\end{figure}

The two-user MIMO IC is considered where transmitter $i$ (denoted as $Tx_i$) has $M_i$ antennas and receiver $i$ (denoted as $Rx_i$) has $N_i$ antennas, respectively, for $i=1,~2$. Such a MIMO IC will be referred to henceforth as the $(M_1,N_1,M_2,N_2)$ MIMO IC. Let the matrix $H_{ij}\in \mathbb{C}^{N_j\times M_i}$ denote the channel between $Tx_i$ and $Rx_j$ with $\|H_{ij}\|_F^2=1$\footnote{There is no loss of generality, since the Frobenius norm of an unnormalized channel matrix can always be absorbed in an SNR or INR (to be soon defined).}. We shall consider a time-invariant or fixed channel where the channel matrices remain fixed for the entire duration of communication. The $(M_1,N_1,M_2,N_2)$ MIMO IC is depicted in Fig. \ref{channel_model_two_user_IC}. At time $t$, $Tx_i$ chooses a vector ${X}_{it}\in \mathbb{C}^{M_i\times 1}$ and sends $\sqrt{P_i}{X}_{it}$ over the channel, where we assume the following average input power constraint at $Tx_i$,

\begin{equation}
\label{power_constraint}
\frac{1}{n}\sum_{t=1}^{n}\textrm{Tr} (Q_{it}) ~\leq ~1,
\end{equation}
for $ i \in \{ 1,2 \}$, where $ Q_{it}=\mathbb{E}({X}_{it}{X}_{it}^{\dagger})$.
Note that in the above power constraint $Q_{it}$'s can depend on the channel matrices.


The received signals at time $t$ can be written as
\begin{eqnarray}
\label{system_eq_two_user_IC2}
Y_{1t}=\sqrt{\rho_{11}}H_{11}X_{1t}+ \sqrt{\rho_{21}}H_{21}X_{2t}+Z_{1t},\\
Y_{2t}= \sqrt{\rho_{22}}H_{22}{X}_{2t}+\sqrt{\rho_{12}} H_{12}X_{1t}+Z_{2t},
\end{eqnarray}
where $Z_{it}\in\mathbb{C}^{N_i\times 1}$ are i.i.d $\mathcal{CN}(\mathbf{0}, I_{N_i})$ across $i$ and $t$, $\rho_{ii}$ and $\rho_{ij}$ represents the signal-to-noise ratio (SNR)\footnote{If the normalized signal vector, $X_i$, has a covariance matrix of $Q_i$, the received signal covariance matrix at $Rx_i$ is $P_iH_{ii}Q_iH_{ii}^\dagger$ and hence the total received signal power is $\textrm{Tr}\left(P_i H_{ii}Q_iH_{ii}^\dagger\right)$ and the corresponding SNR is $ \rho_{ii}=\frac{P_i\textrm{Tr}\left(H_{ii}Q_iH_{ii}^\dagger\right)}{N_i}. $. The INRs of the channel, i.e., $\rho_{ij}$ can be similarly computed.} at receiver $i$ and interference-to-noise ratio (INR) at receiver $j$, respectively for $i \neq j\in \{1,2\}$. In what follows, the MIMO IC with channel matrices, SNRs and INRs as described above will be denoted by $\mathcal{IC}\left(\mathcal{H},\bar{\rho}\right)$, where $\mathcal{H}=\{H_{11},H_{12},H_{21},H_{22}\}$ and $\bar{\rho}=[\rho_{11},\rho_{12},\rho_{21},\rho_{22}]$. The capacity region of $\mathcal{IC}\left(\mathcal{H},\bar{\rho}\right)$ will be denoted by $\mathcal{C}\left(\mathcal{H},\bar{\rho}\right)$ and is defined as follows.

Let us assume that user $i$ transmits information at a rate of $R_i$ to $Rx_i$ using the codebook $\mathcal{C}_{i,n}$ of $n$-length codewords with $|\mathcal{C}_{i,n}|=2^{nR_i}$. Given a message $m_i\in \{1,\cdots , 2^{nR_i}\}$, the corresponding codeword $X_i^n(m_i)\in \mathcal{C}_{i,n}$ must satisfy the power constraint given in \eqref{power_constraint}. From the received signal $Y_i^n$, the receiver obtains an estimate $\hat{m}_i$ of the transmitted message $m_i$ using a decoding function $f_{i,n}$, i.e., $f_{i,n}(Y_i^n)=\hat{m}_i$. Let the average probability of error be denoted by $e_{i,n}=\mathbb{E}\left(\Pr\left(\hat{m}_i\neq m_i\right)\right)$.

A rate pair $(R_1,R_2)$ is achievable if there exists a family of codebooks $\{\mathcal{C}_{i,n}, 1\leq i\leq 2\}_n$ and decoding functions $\{f_{i,n}(.),1\leq i\leq 2\}_n$ such that $\max_i\{e_{i,n}\}$ goes to zero as the block length $n$ goes to infinity. The capacity region $\mathcal{C}(\mathcal{H},\bar{\rho})$ of $\mathcal{IC}\left(\mathcal{H},\bar{\rho}\right)$ is defined as the closure of the set of achievable rate pairs.

\begin{defn}
An achievable rate region is said to be within $(l_1,l_2)$ bits of the capacity region if for any given rate pair $(R_1,R_2)\in \mathcal{C}(\mathcal{H},\bar{\rho})$ the rate pair $((R_1-l_1)^+,(R_2-l_2)^+)$ lies in the achievable region.
\end{defn}


\section{Capacity to within a Constant Gap}
\label{sec_approximate_capacity}
In this section, we characterize the capacity region of the two-user MIMO IC to within a constant number of bits where the constant is independent of SNRs, INRs and the channel matrices. Such a characterization involves establishing a rate region and showing that no rate pair in the capacity region can be further from all the points in the achievable region by more than this constant. Such a characterization of the capacity region will sometimes be referred as the approximate capacity of the channel and the constant as the {\it gap} of approximation. A coding scheme which can achieve a rate region that is within a constant number of bits will be called an approximate capacity optimal or constant-gap-to-capacity optimal coding scheme.

\begin{figure*}
\begin{center}\begin{IEEEeqnarray}{rl}
\label{eq:Rnottt}
\mathcal{R}_o(Q,X_1 ,X_2 ) = \Big\{ (R_l, R_2):
R_1 \leq & h(Y_1|X_2,Q) -h(S_2|X_2,Q), \label{tt1ob}\\
R_1 + R_2 \leq & h(Y_1 |\tilde{U}_1,X_2,Q) + h(Y_2|Q)
-h(S_1|X_1, Q)- h(S_2|X_2,Q),  \\
2R_1 + R_2 \leq  h(Y_1|\tilde{U}_1,X_2,Q) + & h(Y_1|Q) + h(Y_2|\tilde{U}_2,Q)
-h(S_1|X_1, Q) -2h(S_2|X_2,Q),  \\
R_2 \leq & h(Y_2| X_1,Q) -h(S_1|X_1,Q), \\
R_1 + R_2 \leq & h(Y_2 |\tilde{U}_2,X_1,Q) + h(Y_1|Q)
-h(S_1|X_1, Q)- h(S_2|X_2,Q),  \\
R_1 + 2R_2 \leq  h(Y_2| \tilde{U}_2,X_1,Q) + & h(Y_2|Q) + h(Y_1| \tilde{U}_1,Q)
-h(S_2|X_2,Q) -2h(S_1|X_1, Q), \\
R_1 + R_2 \leq & h(Y_1|\tilde{U}_1,Q) + h(Y_2| \tilde{U}_2,Q)
-h(S_1|X_1, Q) -h(S_2|X_2,Q)\Big\}, \label{tt7ob}
\end{IEEEeqnarray}
\end{center}
\hrule
\end{figure*}
\begin{figure*}
\begin{center}
\begin{IEEEeqnarray}{rl}
\label{eq_bound1}
R_1\leq \log \det &\Big[I_{N_1}+\rho_{11} H_{11}H_{11}^{\dagger}\Big];\\
\label{eq_bound2}
R_2\leq \log \det &\Big[I_{N_2}+\rho_{22} H_{22}H_{22}^{\dagger}\Big]; \\
\label{eq_bound3}
R_1+R_2\leq \log \det &\Big[I_{N_2}+\rho_{12} H_{12}H_{12}^{\dagger}+\rho_{22} H_{22}H_{22}^{\dagger}\Big]
 +\log \det \Big[I_{N_1}+\rho_{11} H_{11}K_1H_{11}^{\dagger}\Big];\\
\label{eq_bound4}
R_1+R_2\leq \log \det &\Big[I_{N_1}+\rho_{21} H_{21}H_{21}^{\dagger}+\rho_{11} H_{11}H_{11}^{\dagger}\Big]
 +\log \det \Big[I_{N_2}+\rho_{22} H_{22}K_2H_{22}^{\dagger}\Big];\\
 \label{eq_bound5}
R_1+R_2\leq \log \det &\Big[I_{N_1}+ \rho_{21} H_{21}H_{21}^{\dagger}+  \rho_{11} H_{11} K_1H_{11}^{\dagger}\Big]
+\log \det \Big[I_{N_2}+\rho_{12} H_{12}H_{12}^{\dagger}+\rho_{22} H_{22} K_2H_{22}^{\dagger}\Big];\\
2R_1+R_2\leq \log \det &\Big[I_{N_1}+\rho_{21}  H_{21}H_{21}^{\dagger}+ \rho_{11} H_{11}H_{11}^{\dagger}\Big]+\log\det\Big[I_{N_1}+\rho_{11} H_{11} K_1H_{11}^{\dagger}\Big]+ \nonumber \\
\label{eq_bound6}
& \log \det \Big[I_{N_2}+ \rho_{12}  H_{12}H_{12}^{\dagger}+ \rho_{22} H_{22} K_2H_{22}^{\dagger}\Big];\\
R_1+2R_2\leq \log \det &\Big[I_{N_2}+\rho_{12}  H_{12}H_{12}^{\dagger}+ \rho_{22} H_{22}H_{22}^{\dagger}\Big]+\log\det\Big[I_{N_2}+\rho_{22} H_{22} K_2H_{22}^{\dagger}\Big]+ \nonumber \\
\label{eq_bound7}
&\log \det \Big[I_{N_1}+ \rho_{21} H_{21}H_{21}^{\dagger}+\rho_{11} H_{11} K_1H_{11}^{\dagger}\Big].
\end{IEEEeqnarray}
\end{center}
\hrule
\end{figure*}
In what follows, we shall first obtain a set of explicit upper bounds to the capacity region in terms of the channel matrices. We then give an operational interpretation of these bounds which in turn helps us identify a particular input distribution and linear superposition scheme (by specifying the covariance matrices for the private and public message of each user) leading to a simple HK coding scheme. The achievable rate region of this coding scheme and the corresponding gap to approximate capacity is computed in Section~\ref{subsection_HK_I}. Comparing these set of upper and lower bounds we prove that the two bounds are within $(n_1,n_2)$ bits of each other -- thus proving the constant gap capacity result -- where
\begin{IEEEeqnarray}{rl}
n_i=\hat{m}_{ji}+&\max\Big\{\left(m_{ii}\log(M_i)+m_{ij}\log(M_i+1)\right),\nonumber\\
\label{eq_def_ni}
&~~~~~~~~~~~~~~~~~~~\min\{N_i,M_s\}\log(M_x)\Big\},
\end{IEEEeqnarray}
for~ $1\leq i\neq j\leq 2$, with $M_x=\max\{M_1,M_2\}$, $M_s=(M_1+M_2)$, $m_{ij}$ representing the rank of the matrix $H_{ij}$, and $\hat{m}_{ij}=m_{ij}\log\left(\frac{(M_i+1)}{M_i}\right)$. Note that $m_{ij}\leq \min\{M_i,N_j\}$.

In Section \ref{subsection_HK_II}, an improvement is proposed by allowing the transmitters to select one of three carefully chosen superposition strategies depending on the rate pair to be achieved. It will be shown that the achievable region of this explicit HK coding scheme is within $(n_1^*,n_2^*)$ bits to the capacity region, where
\begin{equation}
\label{eq_def_nis}
n_i^*=\min\{N_i,M_s\}\log(M_x)+\hat{m}_{ji}, ~\textrm{for}~ 1\leq i\neq j\leq 2.
\end{equation}
Note that on a SIMO IC, $n_i^*=1$ (whereas $n_i=2$). 
Finally, in Section \ref{subsection_reciprocity}, we prove the constant gap {\em reciprocity} of the MIMO IC, i.e., the capacity of the two-user MIMO IC does not change by more than a constant number of bits if the roles of the transmitters and receivers are interchanged.

\begin{figure*}
\begin{center}
\begin{IEEEeqnarray}{rl}
  h(Y_i|\tilde{U}_i)\leq h(Y_i^G|\tilde{U}_i^G)
\stackrel{(a)}{=}& \log \det\Big[I_{N_i}+\rho_{ii}H_{ii}{Q}_{i}H_{ii}^{\dagger}+\rho_{ji}H_{ji}{Q}_{j}H_{ji}^{\dagger} \nonumber \\
  &~~~~~~~~~~~~- \rho_{ii}\rho_{ij}H_{ii}{Q}_{i}H_{ij}^{\dagger}\Big[I_{N_j}+\rho_{ij}H_{ij}{Q}_{i}H_{ij}^{\dagger}\Big]^{-1}
H_{ij}{Q}_{i}H_{ii}^{\dagger}\Big]+N_i\log(2\pi e), \nonumber \\
\stackrel{(b)}{=}&\log \det\Big[I_{N_i}+\rho_{ji}H_{ji}{Q}_{j}H_{ji}^{\dagger}+
\rho_{ii}H_{ii}{Q}_{i}^{\frac{1}{2}}\Big[I_{M_i}+\rho_{ij}{Q}_{i}^{\frac{1}{2}}H_{ij}^{\dagger}H_{ij}{Q}_{1}^{\frac{1}{2}}\Big]^{-1}{Q}_{i}^{\frac{1}{2}}H_{ii}^{\dagger}\Big] +N_i\log(2\pi e), \nonumber \\
\stackrel{(c)}{\leq} &  \log \det\Big[I_{N_i}+\rho_{ji} H_{ji} H_{ji}^{\dagger}+
\rho_{ii} H_{ii}\Big[I_{M_i}+\rho_{ij}H_{ij}^{\dagger}H_{ij}\Big]^{-1}H_{ii}^{\dagger}\Big] +N_i\log(2\pi e),
\label{eq:bndyu}
\end{IEEEeqnarray}
\end{center}
\hrule
\end{figure*}
\subsection{An explicit outer bound to the capacity region}
\label{sub_upper_bound}

An outer bound is derived in \cite{TT} for a class of two-user deterministic interference channels which also includes the Gaussian MIMO IC as a special case. In particular, Theorem~$1$ of \cite{TT} when specialized to the Gaussian MIMO IC states that an outer bound to the capacity region is
\begin{equation}
\label{ttouterbound}
\mathcal{R}_o=\cup_{Q,X_1,X_2} \mathcal{R}_o(Q,X_1,X_2).
\end{equation}
where the union is over the timesharing parameter $Q$ and the conditionally independent input vectors $X_i,~1\leq i\leq 2$ given $Q$ taking arbitrary distributions satisfying the power constraint \eqref{power_constraint} and $\mathcal{R}_o(Q,X_1,X_2)$ is given by equations \eqref{tt1ob}-\eqref{tt7ob} at the top of this page in which we have
\begin{IEEEeqnarray}{rl}
Y_{i}=&\sqrt{\rho_{ii}}H_{ii}X_i+\sqrt{\rho_{ji}}H_{ji}X_j+Z_i,~\textrm{for}~i\neq j\in\{1,2\},\nonumber \\
S_i=&\sqrt{\rho_{ij}}H_{ij} X_i+Z_j,
\label{defineyi}
\end{IEEEeqnarray}
with conditionally independent $X_i \in \mathbb{C}^{M_i\times 1}$ (given $Q$) such that $\textrm{Tr}(Q_i) \leq 1$ with $Q_i \triangleq \mathbb{E}(X_i X_i^\dagger|Q)$, $Z_i\sim \mathcal{CN}(0,I_{N_i})$ (with $Z_1$ and $ Z_2$ being independent) and $(\tilde{U}_1,\tilde{U}_2)$ is a conditionally independent copy of $(S_1,S_2)$, so that
\begin{IEEEeqnarray}{l}
\tilde{U}_i=\sqrt{\rho_{ij}}H_{ij}X_i+\tilde{Z}_j,~\forall ~i\neq j\in \{1,2\},
\label{uitilde}
\end{IEEEeqnarray}
and $\tilde{Z}_i\sim \mathcal{CN}(0,I_{N_i})$ for $i=1,2$ and $Z_1$ and $Z_2$ are mutually independent.

We use the above result as a starting point to obtain an explicit outer bound on the capacity region of $\mathcal{IC}(\mathcal{H},\bar{\rho})$, denoted as $\mathcal{R}^u(\mathcal{H},\bar{\rho})$. For economy of notation, we define the matrices
\begin{equation}
\label{eq-defk}
K_i \triangleq \left( I_{M_i}+\rho_{ij} H_{ij}^{\dagger}H_{ij}\right)^{-1} \quad 1\leq i\neq j\leq 2.
\end{equation}

\begin{lemma}[The Outer Bound]
\label{lem_upper_bound}
For a given $\mathcal{H}$ and $\bar{\rho}$ the capacity region, $\mathcal{C}(\mathcal{H},\bar{\rho})$, of the MIMO Gaussian IC with average input power constraint \eqref{power_constraint}, is contained within the set of rate pairs $\mathcal{R}^u(\mathcal{H},\bar{\rho})$, i.e.,
\[\mathcal{C}(\mathcal{H},\bar{\rho})~\subseteq~\mathcal{R}^u(\mathcal{H},\bar{\rho}),\]
where $\mathcal{R}^u(\mathcal{H},\bar{\rho})$ represents the set of rate pairs $(R_1,R_2)$, satisfying equations \eqref{eq_bound1}-\eqref{eq_bound7} for $1\leq i\neq j\leq 2$.
\end{lemma}

\begin{IEEEproof}[Proof of Lemma~\ref{lem_upper_bound}]
The proof is based on a relaxation of the outer bound of \cite{TT} in \eqref{ttouterbound}.
Since conditioning does not increase the average entropy and because the noise vectors at the receivers are independent of the time sharing parameter at the inputs, the rate region $\mathcal{R}_o(Q,X_1,X_2)$ can be outer bounded as in equations~\eqref{eq:UB-1}-\eqref{eq:UB-7}.

\begin{IEEEeqnarray}{rl}
\mathcal{R}_o(Q,X_1 ,X_2& ) \subseteq \Big\{ (R_l, R_2):\nonumber\\
\label{eq:UB-1}
R_1 \leq & h(Y_1|X_2) -h(Z_1),\\
R_1 + R_2 \leq & h(Y_1 |\tilde{U}_1,X_2) + h(Y_2)
-h(Z_2)- h(Z_1),\\
2R_1 + R_2 \leq & h(Y_1|\tilde{U}_1,X_2) + h(Y_1) + h(Y_2|\tilde{U}_2)\nonumber\\
&~~~~~~~~-h(Z_2) -2h(Z_1),\\
R_2 \leq & h(Y_2| X_1) -h(Z_2),\\
R_1 + R_2 \leq & h(Y_2 |\tilde{U}_2,X_1) + h(Y_1)
-h(Z_2)- h(Z_1),\\
R_1 + 2R_2 \leq & h(Y_2| \tilde{U}_2,X_1) + h(Y_2) + h(Y_1| \tilde{U}_1)\nonumber\\
&~~~~~~~~-2h(Z_2) -h(Z_1),\\
\label{eq:UB-7}
R_1 + R_2 \leq & h(Y_1|\tilde{U}_1) + h(Y_2| \tilde{U}_2)
-h(Z_2) -h(Z_1)\Big\},~~~
\end{IEEEeqnarray}

In the rest of the proof, we assume without loss of generality that $\mathbb{E}(X_i)=0$ (since a non-zero mean only contributes to power inefficiency). Using the maximum entropy principle and the fact that $\textrm{Tr}(Q_i) \leq 1 $ implies $Q_i \preceq I_{M_i}$, we have for $i\neq j\in \{1,2\},$
\begin{IEEEeqnarray}{rl}
h(Y_i|X_j) \leq & \log \det\left(I_{N_i}+ \rho_{ii} H_{ii} Q_i H_{ii}^{\dagger}\right)+N_{i}\log(2\pi e), \nonumber \\ \leq & \log \det\left(I_{N_i}+ \rho_{ii} H_{ii} H_{ii}^{\dagger}\right)+N_{i}\log(2\pi e),
\label{eq:bndyx}
\end{IEEEeqnarray}
where the second inequality holds since $Q_i \preceq I_{M_i}$ implies $ I_{N_i}+ \rho_{ii} H_{ii} Q_i H_{ii}^{\dagger} \preceq  I_{N_i}+ \rho_{ii} H_{ii} H_{ii}^{\dagger} $ and because the $ \log \det $ function is monotonically increasing on the cone of p.d. matrices.

Lemma 1 of \cite{thomas} states that given two (complex) zero-mean random vectors of dimensions $t_1$ and $t_2$, respectively, with a joint covariance matrix $K$, the conditional differential entropy of the first one given the second is upper bounded by the conditional differential entropy of a zero-mean Gaussian random vector of dimension $t_1$ given another zero-mean Gaussian random vector of dimension $t_2$ and whose joint covariance is also $K$. Using this result, we have that $ h(Y_i|\tilde{U}_i)\leq h(Y_i^G|\tilde{U}_i^G) $,
where $Y_i^G$ and $\tilde{U}_i^G$ are defined as $Y_i$ and $\tilde{U}_i$ in \eqref{defineyi} and \eqref{uitilde} except with $X_i$ replaced by the Gaussian random vector $X_{i}^G\sim \mathcal{CN}(0,Q_{i})$, with $X_1^G$ and $X_2^G$ mutually independent. Here we have used the standard results that a linear transformation of a Gaussian random vector is Gaussian and a sum of Gaussian random vectors is Gaussian. Hence, we have equation \eqref{eq:bndyu} at the top of the page, where the equality in $(a)$ follows from the fact that $Y_i^G$ given $\tilde{U}_i^G$ is Gaussian and upon evaluating the conditional covariance of $Y_i^G$ given $\tilde{U}_i^G$, the equality in $(b)$ follows from Woodbury's identity and the inequality (c) follows from (i) $\textrm{Tr}(Q_i) \leq 1$ implies $ {Q}_i\preceq I_{M_i} $ (ii) $\log\det(.)$ is a monotonically increasing function on the cone of p.d. matrices and (iii) Lemma~\ref{lem_partial_order_sequence} in Appendix \ref{app:lem_conditional_entropy} with $G_i={Q}_{i}^{\frac{1}{2}}$, $A=H_{ij}^{\dagger}H_{ij}$ and $G_j=I_{M_i}$.

Similarly, it can be shown that
\begin{IEEEeqnarray}{rl}
  h(Y_i|\tilde{U}_i, X_j) \leq &\log \det\left(I_{N_i}+ \rho_{ii} H_{ii}Q_{ij}^{-1}H_{ii}^{\dagger}\right)\nonumber \\
  &~~~~~~~~~~~~~~~~~~~~+N_i\log(2\pi e),
  \label{eq:bndyux}
  \end{IEEEeqnarray}
with $Q_{ij}=(I_{M_i}+\rho_{ij}H_{ij}^{\dagger}H_{ij})$ and the upper bound being the conditional entropy $ h(Y_i|\tilde{U}_i, X_j) $ evaluated for $X_i \sim \mathcal{CN}(0,I_{N_i})$.

Substituting the upper bounds in \eqref{eq:bndyx}, \eqref{eq:bndyu} and \eqref{eq:bndyux} in the bounds of inequalities \eqref{eq:UB-1}-\eqref{eq:UB-7} we obtain $\mathcal{R}^u(\mathcal{H},\bar{\rho})$ as defined in the statement of the lemma, so that
\begin{IEEEeqnarray}{rl}
\mathcal{R}_o(Q,X_1 ,X_2 ) \subseteq \mathcal{R}^u(\mathcal{H},\bar{\rho}).
\end{IEEEeqnarray}
Note that all the seven bounds on the linear combination of rates in $\mathcal{R}_o(Q,X_1,X_2)$ specified in the inequalities \eqref{tt1ob}-\eqref{tt7ob} are simultaneously maximized by independent Gaussian inputs $X_i \sim \mathcal{CN}(0,I_{N_i})$, yielding the region $\mathcal{R}^u(\mathcal{H},\bar{\rho})$. Since this region now does not depend on the input distribution and time sharing parameter, $Q$, we have
\begin{IEEEeqnarray}{rl}
\mathcal{R}_o=\cup_{X_1,X_2,Q}\mathcal{R}_o(Q,X_1 ,X_2 ) \subseteq \mathcal{R}^u(\mathcal{H},\bar{\rho})
\end{IEEEeqnarray}
which concludes the proof.
\end{IEEEproof}


\begin{rem}
\label{rem_incorrect_bound_leveque}
It is noted here that a rate region $\mathcal{R}_{00}(H,G)$ was obtained in \cite{EaOlCv} from the simple computation of the ${\cal R}_0$ region of \cite{TT} for the MIMO Gaussian IC with identity input covariance matrices. It was stated therein that to obtain the (fundamental) diversity-multiplexing tradeoff of the slow fading  MIMO IC it is sufficient to consider the (probability that a rate pair is not in) ${\cal R}_{00}$ region  (see also \cite{Sanjay_Varanasi_IC_DMT} and \cite{Sanjay_Varanasi_ZIC_DMT}). However, the $\mathcal{R}_{00}(H,G)$ as stated in \cite{EaOlCv} needs a correction; specifically, in the third additive terms on bounds for $R_1+2R_2$ and $2 R_1+R_2$, the subscripts of the $G$ matrices should be interchanged. With this correction, $\mathcal{R}_{00}(H,G)$ in \cite{EaOlCv} coincides with the outer bound of Lemma \ref{lem_upper_bound}, as it should. Unlike \cite{EaOlCv} however, Lemma \ref{lem_upper_bound} proves that identity input covariance matrices simultaneously maximize the seven bounds on the rates in \eqref{ttouterbound}, and hence that $\mathcal{R}^u(\mathcal{H},\bar{\rho})$ is an outer bound to the capacity region.
\end{rem}


\subsection{A new inner bound on the capacity region via a simple achievable scheme}
\label{subsection_HK_I}
The achievable region of \cite{TT}, which is specified as a union of sub-regions over all possible input distributions and time-sharing schemes, was shown to be within a constant gap (of $(N_1,N_2)$ bits) of the capacity region. It is not clear from that result as to whether a coding scheme corresponding to a single input distribution, or one that corresponds to time sharing between a few carefully chosen input distributions, can achieve the capacity of the Gaussian MIMO IC within a constant number of bits.
In this section, we develop a simple HK coding scheme corresponding to a single joint distribution of input and auxiliary random variables -- that does not belong to the class of distributions that would be consistent with the prescription of \cite{TT} -- but that nevertheless has the desirable property of having a rate region that is within a constant gap to the outer bound of Lemma \ref{lem_upper_bound}, and hence to the capacity region.

The organization of Section \ref{subsection_HK_I} is as follows. In Section \ref{subsubsec:review}, we briefly review the original HK coding scheme~\cite{Han_Kobayashi} and some recent  developments in \cite{CMG,KH2007},  for the discrete memoryless interference channel (DM-IC) and then apply those results to the Gaussian MIMO IC. This serves to not only introduce notation that makes the ensuing discussions concise but also to explain how this work is related to \cite{Han_Kobayashi,CMG,KH2007}. In Section \ref{subsubsec:interpretation}, we give a novel interpretation for the outer bound of Lemma \ref{lem_upper_bound}. This interpretation is then used as the basis for the specification of a simple achievable scheme in Section \ref{subsubsec:simple}. 
An inner bound on the achievable rate region of the simple HK scheme of this section is then obtained in Section \ref{subsubsec-constgap}. This inner bound is seen to resemble the outer bound of Lemma \ref{lem_upper_bound} from which the constant gap to capacity result is easily deduced.

\subsubsection{A review of HK achievable region and related work}
\label{subsubsec:review}

On a DM-IC with transition probability $P(Y_1,Y_2|X_1,X_2)$, for any set $\mathcal{P}^{\ast}$ of probability distributions $P^*$ of inputs $X_1, X_2$, a time-sharing random variable $Q$ and four auxiliary random variables $U_1,U_2,W_1,W_2$ (defined on arbitrary finite sets) which factor as
\begin{IEEEeqnarray}{l}
\label{eq:input-distribution-DMC}
P^{\ast}(Q,U_1,U_2,W_1,W_2,X_1,X_2)= P(Q) \Pi_{i=1}^2\Big[P(U_i|Q)\nonumber\\
~~~~~~~~~~~~~~~~~~~~~~~~~~~~~~~P(W_i|Q) P(X_i|U_i,W_i,Q)\Big],
\end{IEEEeqnarray}
where $P(X_1|U_1,W_1,Q)$ and $ P(X_2|U_2,W_2,Q)$ are equal to either $0$ or $1$, let
\begin{equation}
\mathcal{R}_{\textrm{HK}}^{o}(P^*)\triangleq \mathcal{R}_{\textrm{HK}}^{(o,1)}(P^*)\cap \mathcal{R}_{\textrm{HK}}^{(o,2)}(P^*) \label{eqhk}
\end{equation}
represent a set of sub-rate 4-tuples defined in what follows. Let $U_i$ and $W_i$ represent the private and common parts of the message to be transmitted by $Tx_i$, referred to henceforth as the private and common message of $Tx_i$ for each $i \in \{1,2\}$. Also, let $r_{iu}$ and $r_{iw}$ represent the information rates carried by $U_i$ and $W_i$, respectively, and let $X_i$ be constructed from $U_i$ and $W_i$ in such a manner that the joint distribution $P^*(Q,U_1,W_1,U_2,W_2,X_1,X_2)\in \mathcal{P}^*$. In \eqref{eqhk}, $\mathcal{R}_{\textrm{HK}}^{(o,i)}(P^*)$ represents the achievable rate region of the three-user multiple-access channel (MAC) in which the transmitters send and the receiver decodes the private and public messages of user $i$ and the private message of user $j\neq i$ to receiver $i$, so that
\begin{subequations}
\label{eq:DMC-sub-rate-region}
\begin{align}
\mathcal{R}_{\textrm{HK}}^{(o,i)}(P^*)=\Big\{(r_{1u},r_{1w},r_{2u},r_{2w}):~~~~~~~~~~~~~~~~~~~\nonumber\\
r_{iu}\leq I(U_i;Y_i|W_i,W_j,Q)\triangleq I_{a_i};\\
\label{eq:DMC-sub-rate-region-1b}
r_{iw}\leq  I(W_i;Y_i|U_i, W_j, Q)\triangleq I_{b_i};\\
\label{eq:DMC-sub-rate-region-1c}
r_{jw}\leq  I(W_j;Y_i|U_i, W_i, Q)\triangleq I_{c_i};\\
(r_{iu}+r_{iw})\leq  I(U_i, W_i;Y_i|W_j, Q)\triangleq I_{d_i};\\
(r_{iu}+r_{jw})\leq  I(U_i, W_j;Y_i|W_i,Q)\triangleq I_{e_i};\\
\label{eq:DMC-sub-rate-region-1f}
(r_{iw}+r_{jw})\leq  I(W_i, W_j;Y_i|U_i, Q)\triangleq I_{f_i};\\
\label{eq:original-HK-rate-region-1g}
(r_{iu}+r_{iw}+r_{jw})\leq  I(U_i, W_i, W_j;Y_i|Q)\triangleq I_{g_i} \Big\}
\end{align}
\end{subequations}
for $ i\neq j\in \{1,2\} $. Further, for a set $\mathcal{S}$ of 4-tuples $(r_{1u},r_{1w},r_{2u},r_{2w})$, define $\Pi(\mathcal{S}) $ as its
two-dimensional projection $ \Pi(\mathcal{S}) \triangleq \{(R_1,R_2): 0\leq R_i\leq (r_{iu}+r_{iw}),1\leq i\leq 2,~ \textrm{for some}~ (r_{1u},r_{1w},r_{2u},r_{2w})\in \mathcal{S}\}$. Then from \cite{Han_Kobayashi} we have the following theorem which essentially states that, for any $(r_{1u},r_{1w},r_{2u},r_{2w})\in \mathcal{R}_{\textrm{HK}}^o(P^*)$, the rate pair $(r_{1u}+r_{1w},r_{2u}+r_{2w})$ is achievable on the DM-IC.
\begin{thm}[HK region \cite{Han_Kobayashi}]\label{thm-ref:HK-original-theorem}
The set
\begin{IEEEeqnarray}{rl}
\mathcal{R}_{\textrm{HK}}^{o}=\Pi\left(\bigcup_{P^*\in \mathcal{P}^{\ast}}\mathcal{R}_{\textrm{HK}}^{o}(P^*)\right)
\end{IEEEeqnarray}
is an achievable region for the DM-IC.
\end{thm}

Thus, for any given $P^*$, Theorem~\ref{thm-ref:HK-original-theorem} not only provides a set of achievable rate pairs of the channel in the form of $\Pi\left(\mathcal{R}_{\textrm{HK}}^o(P^*)\right)$, but also provides the set of 4-tuples from which the rates of their private and public messages can be determined. However, to determine the achievable rate region of the channel, it is necessary to obtain the auxiliary sets $\mathcal{R}_{\textrm{HK}}^{(o,1)}(P^*)$ and $\mathcal{R}_{\textrm{HK}}^{(o,2)}(P^*)$ first. This indirect method can be avoided by using the equivalent description of $\Pi\left(\mathcal{R}_{\textrm{HK}}^o(P^*)\right)$ obtained in Lemma 1 of \cite{CMG}, denoted as $\mathcal{R}_{\textrm{HK}}^e(P^{\ast})$\footnote{The superscript ``o" refers to the {\it original} HK scheme~\cite{Han_Kobayashi} and the superscript ``e" emphasizes that $\mathcal{R}^e_{\textrm{HK}}(P^{\ast})$ is an {\it equivalent} description of $\Pi\left(\mathcal{R}^o_{\textrm{HK}}(P^{\ast})\right)$.}, and stated in the lemma below for easy reference.

\begin{lemma}[Lemma~$1$ in \cite{CMG}]
\label{lem-ref:equivalent-HK-rate-region-1}
For a fixed $P^{\ast}\in \mathcal{P}^{\ast}$, let $\mathcal{R}_{\textrm{HK}}^e(P^{\ast})$ be the set of rate pairs $(R_1,R_2)$ satisfying:
\begin{subequations}\label{eq:compact-rate-region-1}
\begin{align}\label{eq:compact-rate-region-1a}
R_1\leq &I_{d_1};\\
\label{eq:compact-rate-region-1b}
R_1\leq &I_{a_1}+I_{c_2};\\
\label{eq:compact-rate-region-1c}
R_2\leq &I_{d_2};\\
\label{eq:compact-rate-region-1d}
R_2\leq &I_{a_2}+I_{c_1};\\
\label{eq:compact-rate-region-1e}
R_1+R_2\leq &I_{g_2}+I_{a_1};\\
R_1+R_2\leq &I_{g_1}+I_{a_2};\\
R_1+R_2\leq &I_{e_2}+I_{e_2};\\
\label{eq:compact-rate-region-1h}
2R_1+R_2\leq &I_{g_1}+I_{a_1}+I_{e_2};\\
\label{eq:compact-rate-region-1i}
R_1+2R_2\leq &I_{g_2}+I_{a_2}+I_{e_1}.
\end{align}
\end{subequations}
The Han-Kobayashi achievable region $\mathcal{R}_{\textrm{HK}}^{o}$ of Theorem \ref{thm-ref:HK-original-theorem} is also given as $\mathcal{R}_{\textrm{HK}}^e=\cup_{P^{\ast}\in \mathcal{P}^{\ast}} \mathcal{R}^e_{\textrm{HK}}(P^{\ast})$.
\end{lemma}

Evidently, the achievable rate region $\mathcal{R}^e_{\textrm{HK}}(P^{\ast})$ in Lemma~\ref{lem-ref:equivalent-HK-rate-region-1} is now specified directly as a set of rate pairs $(R_1,R_2)$ defined through constraints \eqref{eq:compact-rate-region-1a}-\eqref{eq:compact-rate-region-1i}.


Further, in Theorem~2 of \cite{CMG}, the set of rate pairs $(R_1,R_2)$ constrained by all the bounds of \eqref{eq:compact-rate-region-1} except \eqref{eq:compact-rate-region-1b} and \eqref{eq:compact-rate-region-1d} was defined and called the ``compact" rate region. In the rest of this paper, we shall denote this set with input distribution $P^*$ as $\mathcal{R}^c_{\textrm{HK}}(P^{\ast})$. 
Clearly, $\mathcal{R}^e_{\textrm{HK}}(P^{\ast})\subseteq \mathcal{R}^c_{\textrm{HK}}(P^{\ast})$ and it can be shown that there exist channels $P(Y_1,Y_2|X_1,X_2)$ and input distributions for which $\mathcal{R}^e_{\textrm{HK}}(P^{\ast}) \subset \mathcal{R}^c_{\textrm{HK}}(P^{\ast}) $ (e.g., see Fig. 2 of \cite{CMG} or Fig. \ref{fig:difference_with_timesharing-b}, where the blue dashed line represents $\mathcal{R}^c_{\textrm{HK}}(P^{\ast})$ and the diamond-dotted line represents $\mathcal{R}^e_{\textrm{HK}}(P^{\ast})$). However, it was shown in \cite{CMG} that the union of these two regions over all possible input distributions are equal to each other. Moreover, it was shown in Lemma~2 of \cite{CMG} that $\mathcal{R}^c_{\textrm{HK}}(P^{\ast})$ is contained in the union of three different rate regions $\mathcal{R}^e_{\textrm{HK}}(P_i)$, where $P_i$'s for $1\leq i\leq 3$ are chosen carefully.

The above discussion is summarized in the schematic diagram of Fig. \ref{fig:schematic-summary-1}.

\begin{figure}[h]
\begin{center}
\begin{displaymath}
\xymatrix{
\cap_{i=1}^{2}\mathcal{R}^{(o,i)}_{\textrm{HK}}(P^{\ast})\ar[r]^{\Pi(.)} & \stackrel{\Pi\left(\mathcal{R}^o_{\textrm{HK}}(P^{\ast})\right)}{\scriptstyle = \mathcal{R}^e_{\textrm{HK}}(P^{\ast})}\ar[r]&
\cup_{P^{\ast}\in \mathcal{P}^*}\mathcal{R}^e_{\textrm{HK}}(P^{\ast}) \\
& \nparallel &\parallel \\
& \mathcal{R}^c_{\textrm{HK}}(P^{\ast})  \ar[r] & \cup_{P^{\ast}\in \mathcal{P}^*}\mathcal{R}^c_{\textrm{HK}}(P^{\ast}) }
\end{displaymath}
\caption{Schematic diagram of the relations between the original, equivalent and compact rate regions.}
\label{fig:schematic-summary-1}
\end{center}
\end{figure}

\begin{rem}
\label{rem:equivalence-between-lemma1-and-original-formulation}
Lemma \ref{lem-ref:equivalent-HK-rate-region-1} was proved by showing that for any given $P^*\in\mathcal{P}^*$,
\begin{IEEEeqnarray}{rl}
 \mathcal{R}^e_{\textrm{HK}}(P^{\ast})= &\Pi\left(\mathcal{R}^o_{\textrm{HK}}(P^{\ast})\right)\nonumber\\
 \label{eq:projection-formula}
 = &\Pi\left(\mathcal{R}^{(o,1)}_{\textrm{HK}}(P^{\ast})\cap \mathcal{R}^{(o,2)}_{\textrm{HK}}(P^{\ast})\right).
\end{IEEEeqnarray}
Prior to \cite{CMG}, an equivalent description for $\Pi\left(\mathcal{R}^o_{\textrm{HK}}(P^{\ast})\right)$ was derived in \cite{KH2007} using Fourier-Motzkin elimination on the set of constraints given in \eqref{eq:DMC-sub-rate-region}, with that description having two constraints on $(2R_1+R_2)$ and $(R_1+2R_2)$ in addition to those in Lemma~\ref{lem-ref:equivalent-HK-rate-region-1}. In \cite{CMG}, these additional bounds were shown to be redundant resulting in Lemma~\ref{lem-ref:equivalent-HK-rate-region-1}.
\end{rem}

\begin{rem}
\label{rem:compound-ic-result}
Recently, an alternative to the proof of Theorem 2 of \cite{CMG} was given in \cite{raja_prabhakaran_viswanath}. That proof is based on the fact that a rate region $\mathcal{R}_{\textrm{in}}(P_1^*)$ is achievable by a single input distribution of the form $P_1^*(W_1,X_1,$ $W_2,X_2,Q)=P(Q)P(X_1|Q)P(X_2|Q)P(W_1|X_1,Q)P(W_2|X_2,Q)$, where $\mathcal{R}_{\textrm{in}}(P_1^*)$ is the two-dimensional projection of the set of $4$-tuples $\mathcal{R}_{\textrm{in}}^{(4)}(P_1^*)$, i.e.,
\begin{IEEEeqnarray*}{rl}
    \mathcal{R}_{\textrm{in}}(P_1^*)= &\Pi ( \mathcal{R}_{\textrm{in}}^{(4)}(P_1^*)) = \Big\{(R_1,R_2):R_i=(r_{iu}+r_{iw}),\\
    &~~~~~~~~~~~~~\textrm{and} ~(r_{1u},r_{1w},r_{2u},r_{2w})\in \mathcal{R}_{\textrm{in}}^{(4)}(P_1^*)\Big\},
\end{IEEEeqnarray*}
where $\mathcal{R}_{\textrm{in}}^{(4)}(P_1^*)$ is defined as (see equations (224)-(233) of \cite{raja_prabhakaran_viswanath})
\begin{IEEEeqnarray*}{rl}
\mathcal{R}_{\textrm{in}}^{(4)}(P_1^*)=\Big\{ (r_{1u},r_{1w},r_{2u},r_{2w}): r_{iu}\leq & I_{a_i};\\
(r_{iu}+r_{iw})\leq & I_{d_i};\\
(r_{iu}+r_{jw})\leq & I_{e_i};\\
(r_{iu}+r_{iw}++r_{jw})\leq & I_{g_i} ~\textrm{for}~ i=1,2\Big\}.
\end{IEEEeqnarray*}

While no direct expression for the rate region $\mathcal{R}_{\textrm{in}}(P_1^*)$ was given in \cite{raja_prabhakaran_viswanath}, it was shown in Theorem~D of \cite{KH2007} that $\mathcal{R}_{\textrm{in}}(P_1^*)$ is identical to the rate region $\mathcal{R}_{\textrm{CMG}}(P_1^*)$, specified also in Lemma 4 of \cite{CMG}.
The expression for $\mathcal{R}_{\textrm{CMG}}(P_1^*)$ has two additional constraints than those that define $\mathcal{R}_{\textrm{HK}}^c(P_1^*)$ (e.g., see Lemma~4 and Theorem 2 of \cite{CMG}). In other words, $\mathcal{R}_{\textrm{in}}(P_1^*) \neq \mathcal{R}_{\textrm{HK}}^c(P_1^*)$ for all inputs of the form $P_1^*$. However, it was proved in \cite{CMG} that $\cup_{P_1^{\ast}\in \mathcal{P}_1^*}\mathcal{R}_{\textrm{in}}(P_1^*) = \cup_{P_1^{\ast}\in \mathcal{P}_1^*}\mathcal{R}_{\textrm{HK}}^c(P_1^*)$. These results are represented in the schematic diagram of Fig.
\ref{fig:schematic-second}.

\begin{figure}[h]
\begin{displaymath}
\xymatrix{
\mathcal{R}_{\textrm{in}}^{(4)}(P_1^*)  \ar[r]^{\Pi(.)} &  \mathcal{R}_{\textrm{CMG}}(P_1^{\ast}) \ar[r]&
\cup_{P_1^{\ast}\in \mathcal{P}_1^*}\mathcal{R}_{\textrm{CMG}}(P_1^{\ast}) \\
& \nparallel &\parallel \\
& \mathcal{R}^c_{\textrm{HK}}(P_1^{\ast})  \ar[r] & \cup_{P_1^{\ast}\in \mathcal{P}_1^*}\mathcal{R}^c_{\textrm{HK}}(P_1^{\ast}) }
\end{displaymath}
\caption{Schematic diagram of the relations between $\mathcal{R}_{\textrm{in}}(P_1^*)$, $ \mathcal{R}_{\textrm{CMG}}(P_1^{\ast}) $ and compact rate regions.}
\label{fig:schematic-second}
\end{figure}
\end{rem}

Using standard techniques (cf. Chapter 7 of \cite{Gallager}) these discrete-alphabet results can be applied to the Gaussian IC with continuous alphabets. To distinguish them from each other, the rate regions corresponding to $\mathcal{R}_{\textrm{HK}}^{o}(P^*)$, $\mathcal{R}_{\textrm{HK}}^{(o,i)}(P^*)$, $\mathcal{R}_{\textrm{HK}}^{e}(P^*)$ and $\mathcal{R}_{\textrm{HK}}^c(P^*)$ in the Gaussian IC will be denoted as $\mathcal{R}_{\textrm{HK}}^{G_o}(P^*)$,  $\mathcal{R}_{\textrm{HK}}^{(G_o,i)}(P^*)$, $\mathcal{R}_{\textrm{HK}}^{G_e}(P^*)$ and $\mathcal{R}_{\textrm{HK}}^{G_c}(P^*)$, respectively.

Evidently, both the original description of Theorem~\ref{thm-ref:HK-original-theorem} and the alternative description of Lemma~\ref{lem-ref:equivalent-HK-rate-region-1} of the HK coding scheme are given as a union of an infinite number of sub-regions, each corresponding to a particular input distribution and time sharing strategy. Since a complete characterization of this region is prohibitively complicated, we seek in some sense a single good input distribution and time sharing strategy. Indeed, we provide a novel operational interpretation of the bounds of Lemma~\ref{lem_upper_bound} through which such a good choice of input distribution  becomes apparent, leading to a simple HK coding scheme. Moreover, this simple HK coding scheme has a property of being universally good in that it achieves a rate region that is within a constant number of bits to the set of upper bounds of Lemma~\ref{lem_upper_bound}, independently of SNR and the channel parameters.

\subsubsection{An interpretation of the outer bound of Lemma \ref{lem_upper_bound}}
\label{subsubsec:interpretation}
The first two bounds in $\mathcal{R}^u(\mathcal{H},\bar{\rho})$ come from the rate bound on a point-to-point channel. The first term of the third bound given in \eqref{eq_bound3} (for $j=2$) represents the sum rate upper bound of a two-user MAC having channel matrices $H_{i2}$, for $i=1,2$ and Gaussian input with zero mean and scaled identity matrix as covariance. The second term represents the mutual information on a point-to-point channel whose input covariance matrix is $K_1$ (see (\ref{eq-defk}) for the definition of $K_1$). These terms can be given the following operational interpretation. The entire message of $Tx_2$ has to be decoded at $Rx_2$ and some part of the message of $Tx_1$ might be decoded at $Rx_2$. Let us call this latter message the public message of the first user, denoted as $W_1$, having rate $R_{1w}$. Subsequently, let us denote the remaining part of the first user's message by $U_1$ having rate $R_{1u}$ which will be referred to as the private message of the first user. Thus we have $R_1=R_{1w}+R_{1u}$. Now, with respect to $W_1$ and $X_2$, $Rx_2$ acts as a MAC and thus has the following upper bound
\begin{IEEEeqnarray*}{l}
R_{1w}+R_2\leq \log \det \left(I_{N_2}+\rho_{12} H_{12}H_{12}^{\dagger}+\rho_{22} H_{22}H_{22}^{\dagger}\right).
\end{IEEEeqnarray*}
On the other hand, since $U_1$ has to be decoded at $Rx_1$, it has the following point-to-point channel upper bound
\begin{IEEEeqnarray*}{l}
R_{1u}\leq \log \det \left(I_{N_1}+\rho_{11} H_{11}K_1H_{11}^{\dagger}\right),
\end{IEEEeqnarray*}
where $K_1$ is the covariance matrix of $U_1$. These two bounds together imply the third bound in Lemma~\ref{lem_upper_bound}. The $4^{th}$ bound can also be interpreted similarly just by interchanging the role of transmitters. The first term of the fifth bound can be thought as a bound on the private message of $Tx_1$ and the public message of $Tx_2$ which are to be decoded at $Rx_1$, i.e.,
\begin{IEEEeqnarray*}{l}
R_{1u}+R_{2w}\leq \log \det \left(I_{N_1}+ \rho_{21} H_{21}H_{21}^{\dagger}+  \rho_{11} H_{11} K_1H_{11}^{\dagger}\right),
\end{IEEEeqnarray*}
where the private message has the same covariance matrix as before. Similarly, the second term in the $5^{th }$ bound can be interpreted as an upper bound on $(R_{1w}+R_{2u})$, and together, they imply the fifth bound. The other terms of the remaining bounds can be similarly interpreted.

This interpretation motivates a simple HK scheme, where $Q$ is a deterministic number (no time-sharing), the $i^{th}$ user's message is split into private and public messages, with the private message having an input covariance matrix proportional to $K_i$.

\subsubsection{The simple HK coding scheme}
\label{subsubsec:simple}

\begin{defn}[Input distribution]
\label{def:coding-scheme-input-distribution}
Let the private and public messages of the users be encoded using mutually independent random Gaussian codewords and the overall codeword is the additive superposition of the two, i.e., the transmit signals for any particular channel use can be written as
\begin{equation}
\begin{array}{c}
X_1^g=U_1^g+W_1^g;\\
X_2^g=U_2^g+W_2^g,
\end{array}
\end{equation}
where $U_i^g\sim \mathcal{CN}(\mathbf{0},K_{iu})$ and $W_i^g\sim \mathcal{CN}(\mathbf{0},K_{iw})$, represent symbols of the codewords of the private and public messages of user $i$, respectively, and
\begin{IEEEeqnarray}{rl}
K_{iu}(\mathcal{H})\triangleq & \mathbb{E}(U_i^g U_i^{g \dagger}) \nonumber\\
\label{eq_power_split1}
=& \frac{K_i}{M_i}=\frac{1}{M_i}\left(I_{M_i}+\rho_{ij} H_{ij}^{\dagger}H_{ij}\right)^{-1};\\
\label{eq_power_split2}
K_{iw}(\mathcal{H})\triangleq & \mathbb{E}(W_i^g W_i^{g \dagger})=  \frac{1}{M_i}\left(I_{M_i}-K_i\right).
\end{IEEEeqnarray}
The scaling by $\frac{1}{M_i}$ is required to satisfy the power constraint \eqref{power_constraint}. In the sequel, we shall refer to such a superposition coding scheme -- where the covariance matrices of the private and public messages of user $i$ are given by $K_{iu}$ and $K_{iw}$ -- as the $\mathcal{HK}\left(\{K_{iu},K_{iw},K_{iu},K_{iw}\}\right)$ scheme. In particular, when $K_{iu}$ and $K_{iw}$ are as in equation \eqref{eq_power_split1} and \eqref{eq_power_split2}, respectively, the coding scheme will be denoted as $\mathcal{HK}^{(s)}$ (the superscript $s$ stands for ``simple"). Let us denote the distribution of the random variables defined above as $P_s(U_1^g,W_1^g,X_1,U_2^g,W_2^g,X_2)$. Clearly, $P_s(U_1^g,W_1^g,X_1^g,U_2^g,W_2^g,X_2^g)\in\mathcal{P}^{\ast}$ (with no time-sharing).
\end{defn}

\begin{rem}
\label{rem:interference-at-noise-floor}
The above choice ensures that the private message of user $i$, the covariance of the contribution of which at $Rx_j$ (namely, $\sqrt{\rho_{ij}}H_{ij}U_i^g$), is given by
\begin{IEEEeqnarray*}{rl}
\rho_{ij}H_{ij}K_{iu}H_{ij}^\dagger = &\frac{\rho_{ij}}{M_i}H_{ij}\left(I_{M_i}+\rho_{ij} H_{ij}^{\dagger}H_{ij}\right)^{-1}H_{ij}^\dagger \\ \preceq & I_{M_i},
\end{IEEEeqnarray*}
and thus {\em reaches the unintended receiver below the noise floor}. Thus the simple achievable scheme here when specialized to the SISO IC embodies the key principle in the achievable scheme of \cite{ETW1}. It is of course applicable much more generally to MIMO ICs and cannot be as such inferred in its general form from just that principle alone. Indeed, we will also soon see that even when specialized to the SISO IC the details of the power split between the public and private messages resulting from this work are different from those in \cite{ETW1}.

\begin{figure*}
\begin{center}
\begin{IEEEeqnarray}{rl}
\label{eq:Is-for-Gaussian-input-1}
I_{a_1}^g \triangleq I(X_1^g;Y_1^g|W_1^g,W_2^g) = & \log\det \Big[I_{N_1}+\rho_{11}H_{11}K_{1u}H_{11}^{\dagger}+\rho_{21}H_{21}K_{2u}H_{21}^{\dagger}\Big]-\tau_{21};\\
\label{eq:Is-for-Gaussian-input-2}
I_{b_1}^g \triangleq I(W_1^g,;Y_1^g|W_2^g, U_1^g) = & \log\det \Big[I_{N_1}+\rho_{11}H_{11}K_{1w}H_{11}^{\dagger}+\rho_{21}H_{21}K_{2u}H_{21}^{\dagger}\Big]-\tau_{21};\\
\label{eq:Is-for-Gaussian-input-3}
I_{c_1}^g \triangleq I(W_2^g;Y_1^g|X_1^g) = &\log\det\Big[I_{N_1}+M_2^{-1}\rho_{21} H_{21}H_{21}^{\dagger}\Big]-\tau_{21};\\
\label{eq:Is-for-Gaussian-input-4}
I_{d_1}^g \triangleq I(X_1^g;Y_1^g|W_2^g) = &\log\det\Big[I_{N_1}+M_1^{-1}\rho_{11} H_{11}H_{11}^{\dagger}+\rho_{21}H_{21}K_{2u}H_{21}^\dagger\Big]-\tau_{21};\\
\label{eq:Is-for-Gaussian-input-5}
I_{e_1}^g \triangleq I(X_1^g,W_2^g;Y_1^g|W_1^g) = &\log\det\Big[I_{N_1}+ M_2^{-1}\rho_{21} H_{21}H_{21}^{\dagger}+ \rho_{11} H_{11}K_{1u}H_{11}^{\dagger}\Big]-\tau_{21};\\
\label{eq:Is-for-Gaussian-input-6}
I_{f_1}^g \triangleq I(W_1^g,W_2^g;Y_1^g|U_1^g) = & \log\det\Big[I_{N_1}+M_2^{-1}\rho_{21} H_{21}H_{21}^{\dagger}+\rho_{11} H_{11}K_{1w}H_{11}^{\dagger}\Big]-\tau_{21};\\
\label{eq:Is-for-Gaussian-input-7}
I_{g_1}^g \triangleq I(X_1^g,W_2^g;Y_1^g) = & \log\det\Big[I_{N_1}+M_2^{-1}\rho_{21} H_{21}H_{21}^{\dagger}+M_1^{-1}\rho_{11} H_{11}H_{11}^{\dagger}\Big]-\tau_{21},
\end{IEEEeqnarray}
\end{center}
\hrule
\end{figure*}

\begin{figure}[htp]
  \begin{center}
    \includegraphics[height=6cm, width=8cm]{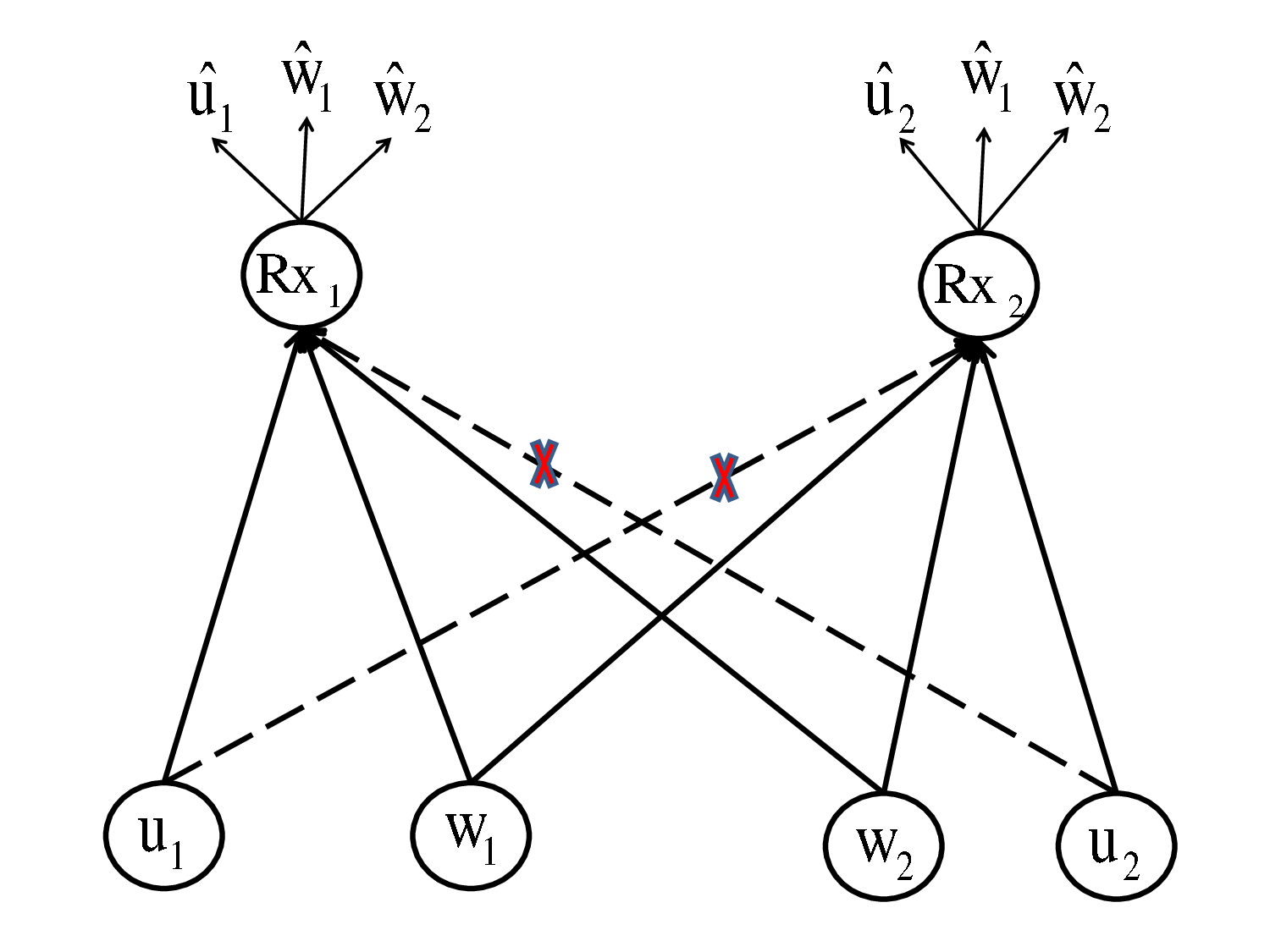}
  \end{center}
\caption{The equivalent virtual channel for the simple HK coding scheme. }
 \label{fig_simple_coding_scheme}
\end{figure}

The $\mathcal{HK}\left(\{K_{1u},K_{1w},K_{2u},K_{2w}\}\right)$ coding scheme thus effectively divides each user into two virtual users as shown in Fig. \ref{fig_simple_coding_scheme}. Note that the interference links from the first virtual user to $Rx_2$ and the fourth virtual user to $Rx_1$ are made very weak so that any signal along those links always reaches the receivers below the noise floor. As shown in the figure, the channel can be thought as two interfering MACs where $Rx_i$ jointly decodes $U_i^g$, $W_i^g$ and $W_{j\neq i}^{g}$, treating $U_j^g$ as noise for $1\leq i\neq j\leq 2$.
\end{rem}

Applying Theorem~\ref{thm-ref:HK-original-theorem} and Lemma~\ref{lem-ref:equivalent-HK-rate-region-1} to the Gaussian case and evaluating for the distribution $P_s(.)$ of Definition~\ref{def:coding-scheme-input-distribution}, we get the following achievable region for the two-user MIMO Gaussian IC.

\begin{lemma}
\label{lem:equivalent-gaussian-rate-region}
On a two-user Gaussian MIMO IC, the simple $\mathcal{HK}\left(\{K_{1u},K_{1w},K_{2u},K_{2w}\}\right)$ coding scheme achieves the rate region, $\mathcal{R}^{G_e}_{\textrm{HK}}(P_s)$, which is a set of rate pairs $(R_1,R_2)$ where $R_i$'s satisfy the bounds \eqref{eq:compact-rate-region-1} of Lemma \ref{lem-ref:equivalent-HK-rate-region-1} so that
\begin{subequations}\label{eq:compact-gaussian-rate-region-1}
\begin{align}\label{eq:compact-gaussian-rate-region-1a}
R_1\leq &I_{d_1}^g;\\
\label{eq:compact-gaussian-rate-region-1b}
R_1\leq &I_{a_1}^g+I_{c_2}^g;\\
\label{eq:compact-gaussian-rate-region-1c}
R_2\leq &I_{d_2}^g;\\
\label{eq:compact-gaussian-rate-region-1d}
R_2\leq &I_{a_2}^g+I_{c_1}^g;\\
\label{eq:compact-gaussian-rate-region-1e}
R_1+R_2\leq &I_{g_2}^g+I_{a_1}^g;\\
R_1+R_2\leq &I_{g_1}^g+I_{a_2}^g;\\
R_1+R_2\leq &I_{e_1}^g+I_{e_2}^g;\\
\label{eq:compact-gaussian-rate-region-1h}
2R_1+R_2\leq &I_{g_1}^g+I_{a_1}^g+I_{e_2}^g;\\
\label{eq:compact-gaussian-rate-region-1i}
R_1+2R_2\leq &I_{g_2}^g+I_{a_2}^g+I_{e_1}^g,
\end{align}
\end{subequations}
where $I_{w_i}^g $'s are equal to the mutual information terms $I_{w_i}$'s defined in \eqref{eq:DMC-sub-rate-region} (with $w$ denoting one of $\{a,b,c,d,e,f,g\}$) evaluated for the input specified in Definition~\ref{def:coding-scheme-input-distribution} and are given in \eqref{eq:Is-for-Gaussian-input-1}-\eqref{eq:Is-for-Gaussian-input-7}.
Further (see Fig. \ref{fig:schematic-summary-1}),
\begin{IEEEeqnarray*}{rl}
    \mathcal{R}^{G_e}_{\textrm{HK}}(P_s)=&\Pi\left(\mathcal{R}^{G_o}_{\textrm{HK}}(P_s)\right)\\
    =&\Pi\left(\mathcal{R}^{(G_o,1)}_{\textrm{HK}}(P_s)\cap \mathcal{R}^{(G_o,2)}_{\textrm{HK}}(P_s)\right),
\end{IEEEeqnarray*}
where $ \mathcal{R}_{\textrm{HK}}^{(G_o,i)}(P_s) $ is as specified by \eqref{eq:DMC-sub-rate-region} with the achievable rate bounds given in terms of the mutual information terms $I_{w_i}$'s therein again replaced by their corresponding $I_{w_i}^g $'s where $\tau_{ij}=\log\det(I_{N_j}+\rho_{ij}H_{ij}K_{iu}H_{ij}^\dagger)$ for $i\neq j \in \{1,2\}$ and $I_{a_2}^g$ through $I_{g_2}^g $ are obtained by swapping the indexes $1$ and $2$ in the above set of equations, where $K_{iu}$ and $K_{iw}$ are given by \eqref{eq_power_split1} and \eqref{eq_power_split2}, respectively, for $1\leq i\leq 2$.
\end{lemma}
\begin{IEEEproof}
The rate bounds for $\mathcal{R}^{G_e}_{\textrm{HK}}(P_s)$ and $ \mathcal{R}_{\textrm{HK}}^{(G_o,i)}(P_s)$ as specified in this lemma result from a simple evaluation of the DM-IC result of Lemma~\ref{lem-ref:equivalent-HK-rate-region-1} and Theorem~\ref{thm-ref:HK-original-theorem} to the MIMO Gaussian IC. Equations \eqref{eq:Is-for-Gaussian-input-1}-\eqref{eq:Is-for-Gaussian-input-7} are obtained by evaluating the different mutual information terms in \eqref{eq:DMC-sub-rate-region} for the distribution $P_s(U_1^g,W_1^g,X_1,U_2^g,W_2^g,X_2)$ of Definition~\ref{def:coding-scheme-input-distribution}.
\end{IEEEproof}

\begin{ex}
\label{ex:achievable-region-of-simple-HK-scheme}
Since the bounds of Lemma \ref{lem_upper_bound} and Lemma \ref{lem:equivalent-gaussian-rate-region} are explicit, one can compute them. Consider a two-user Gaussian $(2,3,2,2)$ IC with $\bar{\rho}=[20,~8,~12,~20]$ dB and the channel matrices given as
\begin{IEEEeqnarray*}{rl}
H_{11}=& \left[\begin{array}{cc}
              1.1975 - 0.4385i & -0.0902 + 0.1895i\\
   0.3234 - 1.3614i &  0.1330 - 0.2564i\\
   0.7546 - 1.0080i & -0.3205 - 0.6958i \end{array}
\right]; \\
H_{21}=& \left[\begin{array}{cc}
              0.3816 - 0.8508i  & 0.4450 - 0.4386i\\
  -0.4892 - 0.2179i & -0.5346 - 0.1519i\\
   0.7665 - 1.0875i &  0.1689 + 0.7651i
             \end{array}
\right];\\
H_{12}=& \left[\begin{array}{cc}
           0.9652 - 0.8085i & -0.3033 + 0.0055i\\
   0.6130 + 1.4479i &  0.6872 + 0.5280i
             \end{array}
\right];\\
H_{22}=&\left[\begin{array}{cc}
              -0.1209 - 0.4575i & -0.0040 + 0.0921i\\
  -0.5730 + 1.1118i & -0.8223 - 0.5687i
             \end{array}
\right];
\end{IEEEeqnarray*}
The explicit outer and inner bounds for this channel are shown in Fig. \ref{fig:achievable-region-simple-HK-scheme}.

\begin{figure*}
\begin{center}
\begin{IEEEeqnarray*}{rl}
R_1\leq \Big[\log \det &\Big[I_{N_1}+\rho_{11} H_{11}H_{11}^{\dagger}\Big]-n_1\Big]^+; \label{eq:rate-region-within-ni-bits1}\\
R_2\leq \Big[\log \det &\Big[I_{N_2}+\rho_{22} H_{22}H_{22}^{\dagger}\Big]-n_2\Big]^+; \label{eq:rate-region-within-ni-bits2} \\
R_1+R_2\leq \Big[\log \det &\Big[I_{N_2}+\rho_{12} H_{12}H_{12}^{\dagger}+\rho_{22} H_{22}H_{22}^{\dagger}\Big]+\log \det \Big[I_{N_1}+\rho_{11} H_{11}K_{1}H_{11}^{\dagger}\Big]-(n_1^*+n_2^*)\Big]^+; \label{eq:rate-region-within-ni-bits3} \\
R_1+R_2\leq \Big[\log \det &\Big[I_{N_1}+\rho_{21} H_{21}H_{21}^{\dagger}+\rho_{11} H_{11}H_{11}^{\dagger}\Big]+\log \det \Big[I_{N_2}+\rho_{22} H_{22}K_{2}H_{22}^{\dagger}\Big]-(n_1^*+n_2^*)\Big]^+; \label{eq:rate-region-within-ni-bits4} \\
R_1+R_2\leq \Big[\log \det &\Big[I_{N_1}+ \rho_{21} H_{21}H_{21}^{\dagger}+ \rho_{11} H_{11}K_{1}H_{11}^{\dagger}\Big]+\log \det \Big[I_{N_2}+\rho_{12} H_{12}H_{12}^{\dagger}+\rho_{22} H_{22} K_{2} H_{22}^{\dagger}\Big] -(n_1^*+n_2^*)\Big]^+; \label{eq:rate-region-within-ni-bits5} \\
2R_1+R_2\leq \Big[\log \det & \Big[I_{N_1}+\rho_{21}  H_{21}H_{21}^{\dagger} + \rho_{11} H_{11}H_{11}^{\dagger}\Big] + \log\det \Big[I_{N_1}+\rho_{11} H_{11} K_{1}H_{11}^{\dagger}\Big] \nonumber\\
&~~~~~~~~~~+ \log \det  \Big[I_{N_2}+ \rho_{12}  H_{12}H_{12}^{\dagger}+\rho_{22} H_{22} K_{2}H_{22}^{\dagger}\Big]-(2n_1^*+n_2^*)\Big]^+; \label{eq:rate-region-within-ni-bits6} \\
R_1+2R_2\leq \Big[\log \det &\Big[I_{N_2}+\rho_{12}  H_{12}H_{12}^{\dagger}+ \rho_{22} H_{22}H_{22}^{\dagger}\Big]  +\log\det \Big[I_{N_2}+ \rho_{22} H_{22} K_{2}H_{22}^{\dagger}\Big] \nonumber\\
\label{eq:rate-region-within-ni-bits7}
&~~~~~~~~~~+ \log \det  \Big[I_{N_1}+ \rho_{21} H_{21}H_{21}^{\dagger}+\rho_{11} H_{11} K_1 H_{11}^{\dagger}\Big]-(n_1^*+2n_2^*)\Big]^+ .
\end{IEEEeqnarray*}
\end{center}
\hrule
\end{figure*}

\begin{figure}[htp]
  \begin{center}
    \includegraphics[height=6cm, width=8cm]{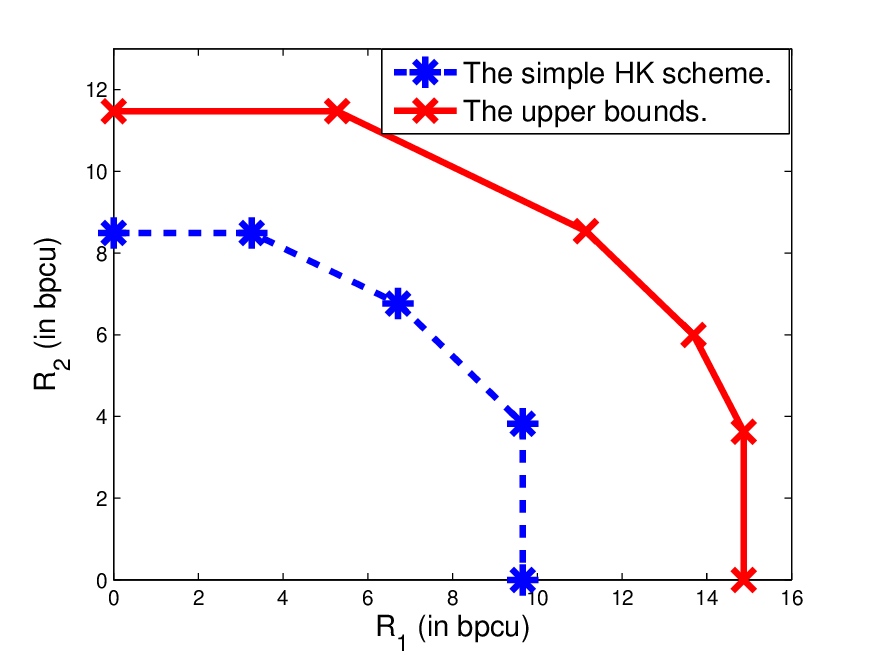}
   \end{center}
\caption{An achievable rate region of the simple HK scheme.}
\label{fig:achievable-region-simple-HK-scheme}
\end{figure}
\end{ex}

\begin{rem}
\label{subsubsec:rate-split}
To achieve a rate pair $(R_1,R_2)\in \mathcal{R}_{\textrm{HK}}^{G_e}(P_s)$, the rates of the private and public messages, namely $r_{iu}$ and $r_{iw}$, have to be chosen (with $ r_{iu}+r_{iw}=R_i,~\forall~i~\in \{1, 2\} $) such that both public and the intended private messages can be decoded at each receiver with arbitrarily reliability. From the second part of Lemma \ref{lem:equivalent-gaussian-rate-region}, such a 4-tuple of $(r_{1u},r_{1w},r_{2u},r_{2w})$ exists in $\mathcal{R}^{G_o}_{\textrm{HK}}(P_s)$ (since $\mathcal{R}^{G_e}_{\textrm{HK}}(P_s)=\Pi\left(\mathcal{R}^{G_o}_{\textrm{HK}}(P_s)\right)$).


\end{rem}

\subsubsection{The constant gap result for the simple HK scheme}
\label{subsubsec-constgap}

It is not unreasonable to imagine that the gap between the boundaries of the achievable rate region and the set $\mathcal{R}^u(\mathcal{H},\bar{\rho})$ can behave arbitrarily, including becoming unbounded sometimes, as a function of the channel matrices. However, in what follows we shall show that this gap remains bounded and cannot be larger than a constant which is independent of the SNR, INR or the channel coefficients. This fact will be proved by showing that $\mathcal{R}^{G_e}_{\textrm{HK}}(P_s)$ contains a subset which is within a constant gap to the set of upper bounds. The following lemma specifies this subset.

\begin{lemma}
\label{lem:rate-region-within-ni-bits}
The achievable rate region $\mathcal{R}^{G_e}_{\textrm{HK}}(P_s) $ contains the rate region $ \mathcal{R}_a(\mathcal{H},\bar{\rho})$ (which is thus also achievable by $\mathcal{HK}(\{K_{1u},K_{1w},K_{1u},K_{1w}\})$), which is a polygon of non-negative rate pairs satisfying equations \eqref{eq:rate-region-within-ni-bits1}-\eqref{eq:rate-region-within-ni-bits7} at the top of the page.
\end{lemma}
\begin{IEEEproof}[Proof]
The proof is given in Appendix \ref{pf:lem:rate-region-within-ni-bits}. 
\end{IEEEproof}

Note that each bound of Lemma~\ref{lem:rate-region-within-ni-bits} differs from the corresponding bound in Lemma~\ref{lem_upper_bound} only by a constant, from which we get the following constant gap to capacity result.

\begin{thm}
\label{thm_approximate_capacity}
The rate region $\mathcal{R}_a\left(\mathcal{H},\bar{\rho}\right)$, which is contained in the rate region achievable by the simple HK scheme $\mathcal{HK}\left(\{K_{1u}, K_{1w}, K_{2u}, K_{2w}\}\right)$, is within $(n_1,n_2)$ bits to the capacity region of the Gaussian MIMO IC, where $n_i$ is given by \eqref{eq_def_ni}.
\end{thm}
\begin{IEEEproof}
The proof is given in Appendix~\ref{app:gap-to-outerbound}.
\end{IEEEproof}


\begin{rem}[MMSE interpretation]
\label{rem:mmse-interpretation}
The form of the covariances of the private messages (given in \eqref{eq_power_split1}) in the $\mathcal{HK}\left(\{K_{1u}, K_{1w}, K_{2u}, K_{2w}\}\right)$ scheme is reminiscent of the error covariance matrix in minimum mean-square error (MMSE) estimation in linear Gaussian models. In what follows, we explore this connection. As indicated by the simple HK scheme, assume that $X_i\sim \mathcal{CN}(0,\frac{1}{M_i}I_{M_i})$, for $i=1,2$. Motivated by the choice in \cite{TT} of the auxiliary (common message) random variables $ (\tilde{U}_1, \tilde{U}_2)$ (denoted as $(U_1, U_2)$ in \cite{TT}) 
be a conditionally independent copy of $(S_1, S_2)$ conditioned on the inputs $(X_1, X_2)$, we let $\tilde{U}_i=\sqrt{\rho_{ij}}H_{ij}X_i+\tilde{Z}_j$. Hence, $(X_i, \tilde{U}_i)$ are jointly Gaussian. Now, let the MMSE estimate of $X_i$ based on $\tilde{U}_i$ be denoted as $\hat{X}_i$ and the estimation error as $E_i$, so that $ X_i=\hat{X_i}+E_i $. It is well known that $\hat{X}_i$ and $E_i$ are jointly Gaussian and independent. 
Moreover, from Theorem~12.1 in \cite{kay} we have
\begin{equation}
\label{error-cov}
E_i\sim \mathcal{CN}\left(0, \frac{1}{M_i}\left(I_{M_i}+\frac{\rho_{ij}}{M_i}H_{ij}^\dagger H_{ij}\right)^{-1}\triangleq \tilde{K}_{iu}\right).
\end{equation}
Denoting the covariance matrix of $E_i$ by $K_{E_i}$, we get the covariance matrix for $\hat{X}_i$, i.e., $\hat{X}_i\sim \mathcal{CN}(0,\frac{1}{M_i}I_{M_i}-K_{E_i})$.
Now, interpret $\hat{X}_i$ as the common message at transmitter $i$, i.e., $W_i=\hat{X_i}$ with the private message $U_i=X_i-\hat{X_i} = E_i$, so that $U_i$ and $W_i$ are independent Gaussian vectors. The resulting $\mathcal{HK}\left(\{\tilde{K}_{1u}, \tilde{K}_{1w}, \tilde{K}_{2u}, \tilde{K}_{2w}\}\right)$ superposition scheme (with $\tilde{K}_{iw} = \frac{1}{M_i}I_{M_i}-K_{E_i}$) bears a resemblance to the $\mathcal{HK}\left(\{K_{1u}, K_{1w}, K_{2u}, K_{2w}\}\right)$ scheme. However, there is a subtle but important difference between the two. In particular, $ \tilde{K}_{iu}$ in \eqref{error-cov} looks similar to $K_{iu}$ in \eqref{eq_power_split1}, but the two matrices are not identical. Hence, starting with the assumption that $X_i\sim \mathcal{CN}(0,\frac{1}{M_i}I_{M_i})$ as dictated by the $\mathcal{HK}\left(\{K_{1u}, K_{1w}, K_{2u}, K_{2w}\}\right)$ scheme, adopting the choice of auxiliary random variables from \cite{TT}, and then specifying the private and public message covariances in a linear superposition scheme using the above MMSE based reasoning yields a simple HK scheme that is not identical to that given in Definition \ref{def:coding-scheme-input-distribution}.

Nevertheless, it can be proved 
that this new MMSE based public-private covariance splitting strategy of $\mathcal{HK}\left(\{\tilde{K}_{1u}, \tilde{K}_{1w}, \tilde{K}_{2u}, \tilde{K}_{2w}\}\right)$ yields an achievable rate region that is also within a constant gap to the set of upper bounds of Lemma \ref{lem_upper_bound}. However, this gap is strictly larger than that found in Theorem~\ref{thm_approximate_capacity}.
In particular, replacing $K_{iu}$ in the bounds for the achievable rate region by $\tilde{K}_{iu}$, we see that the upper bound in \eqref{eq_pf_achieve_lemma_t2} in Appendix~\ref{pf:lem:rate-region-within-ni-bits} becomes $m_{ij}$ instead of $\hat{m}_{ij}$. When we substitute this bound in the subsequent bounding steps in Appendix~\ref{pf:lem:rate-region-within-ni-bits} we see that all the $\hat{m}_{ij}$'s get replaced by $m_{ij}$. Consequently, the gap to the upper bound becomes
\begin{IEEEeqnarray*}{rl}
\tilde{n}_i= \max\Big\{&\left(m_{ii}\log(M_i)+m_{ij}\log(M_i+1)\right),\\
&~~~~~~~~~~~~\min\{N_i,M_s\}\log(M_x)\Big\}+{m}_{ji},
\end{IEEEeqnarray*}
for $i\neq j\in \{1,2\}$. Since $m_{ji} \geq \hat{m}_{ji} $, the above gap is larger than $n_i$. The difference between the two gaps
\begin{equation*}
\tilde{n}_i-n_i=m_{ji}\log\left(\frac{2M_j}{M_j+1}\right),
\end{equation*}
for $~i\neq j\in \{1,2\},$ in fact increases with $m_{ji}$ when $M_j > 1$.
\end{rem}

\subsection{An explicit coding scheme achieves a smaller gap}
\label{subsection_HK_II}

As mentioned previously in Section \ref{subsubsec:review} following Lemma \ref{lem-ref:equivalent-HK-rate-region-1}, the rate pairs in $ \mathcal{R}_{\textrm{HK}}^{G_c}(P_s)$, the compact rate region  corresponding to $P_s(.) $ of Definition \ref{def:coding-scheme-input-distribution}, is the set of non-negative rate pairs $(R_1, R_2)$ that satisfy the set of constraints in \eqref{eq:compact-gaussian-rate-region-1} except the two inequalities \eqref{eq:compact-gaussian-rate-region-1b} and  \eqref{eq:compact-gaussian-rate-region-1d}.


Consider the linear Gaussian superposition HK scheme $\mathcal{HK}(\{\frac{1}{M_1}I_{M_1},\mathbf{0},K_{2u},K_{2w}\})$ in which $Tx_1$'s message  is entirely private whereas $Tx_2$ splits its message into private and public messages with covariance assignments that are the same as for $Tx_2$ in the
$\mathcal{HK}^{(s)}$ scheme (i.e., $X_1^g=U_1^g $ and $X_2^g = U_2^g+W_2^g$). We denote this scheme as the
$\mathcal{HK}^{(s_1)}$ scheme. Similarly, consider the $\mathcal{HK}(\{K_{1u},K_{1w},\frac{1}{M_2}I_{M_2},\mathbf{0}\})$ scheme in which $Tx_2$'s entire message is private with $Tx_1$ splitting its message into private and public messages with covariance assignments being the same as for $Tx_1$ in the $\mathcal{HK}^{(s)}$ scheme. This latter scheme is referred to as the $\mathcal{HK}^{(s_2)}$ scheme. Also, let $ \mathcal{R}_{\textrm{HK}}^{G_e}(P_{s_1})$ and $ \mathcal{R}_{\textrm{HK}}^{G_e}(P_{s_2})$ denote the corresponding equivalent HK rate regions, respectively. The distributions $P_{s_1}, P_{s_2} \in {\cal P}^{\ast} $ denote the distributions of inputs and auxiliary random variables associated with the simple $\mathcal{HK}^{(s_1)}$ and $\mathcal{HK}^{(s_2)}$ schemes.

\begin{defn}[The explicit HK scheme] For any rate pair $(R_1,R_2) \in \mathcal{R}_{\textrm{HK}}^{G_e}(P_s) \cup \mathcal{R}_{\textrm{HK}}^{G_e}(P_{s_1}) \cup \mathcal{R}_{\textrm{HK}}^{G_e}(P_{s_2})$, there exists at least one distribution $P \in \{P_s, P_{s_1}, P_{s_2}\} $ such that $(R_1,R_2) \in \mathcal{R}_{\textrm{HK}}^{G_e}(P) $. For this rate pair, select the coding scheme from $\mathcal{HK}^{(s)}$ or $\mathcal{HK}^{(s_1)}$ or $\mathcal{HK}^{(s_2)}$ that corresponds to the distribution $P$. The resulting explicit scheme is denoted as $\widetilde{\mathcal{HK}}$ (and has an achievable rate region $\mathcal{R}_{\textrm{HK}}^{G_e}(P_s) \cup \mathcal{R}_{\textrm{HK}}^{G_e}(P_{s_1}) \cup \mathcal{R}_{\textrm{HK}}^{G_e}(P_{s_2})$).
\end{defn}
\begin{figure*}
  \begin{center}
    \subfigure[Achievable rate regions of the component schemes $\mathcal{HK}^{(s)}, \mathcal{HK}^{(s_1)}$, and $\mathcal{HK}^{(s_2)}$.]{\label{fig:difference_with_timesharing-a}\includegraphics[scale=.5]{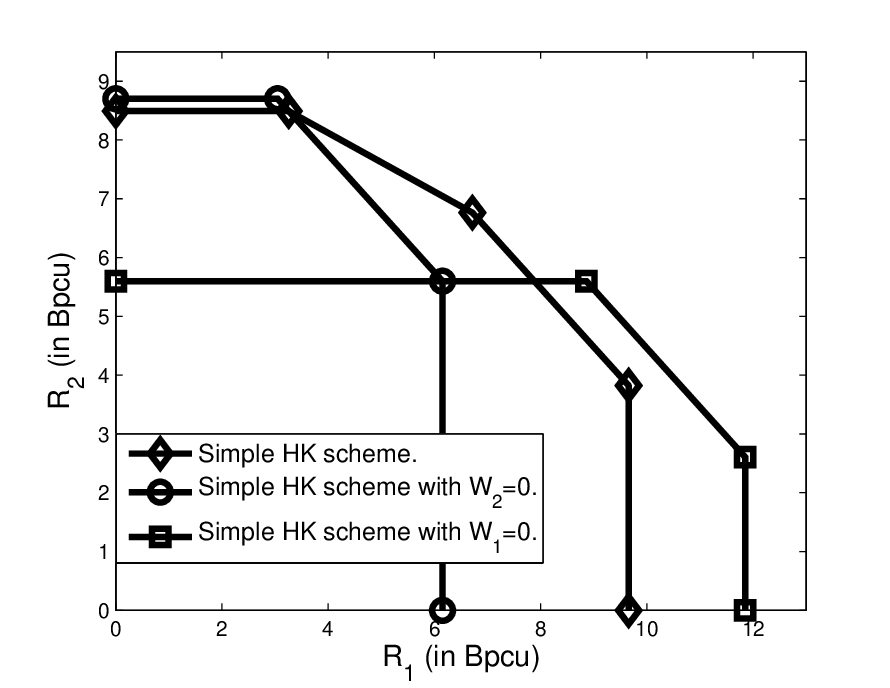}}
    \subfigure[The compact rate region $\mathcal{R}_{\textrm{HK}}^{G_c}(P_s) $ is strictly contained in that achievable by the explicit scheme. Time sharing strictly enlarges the rate region of the explicit scheme.]{\label{fig:difference_with_timesharing-b}\includegraphics[scale=0.5]{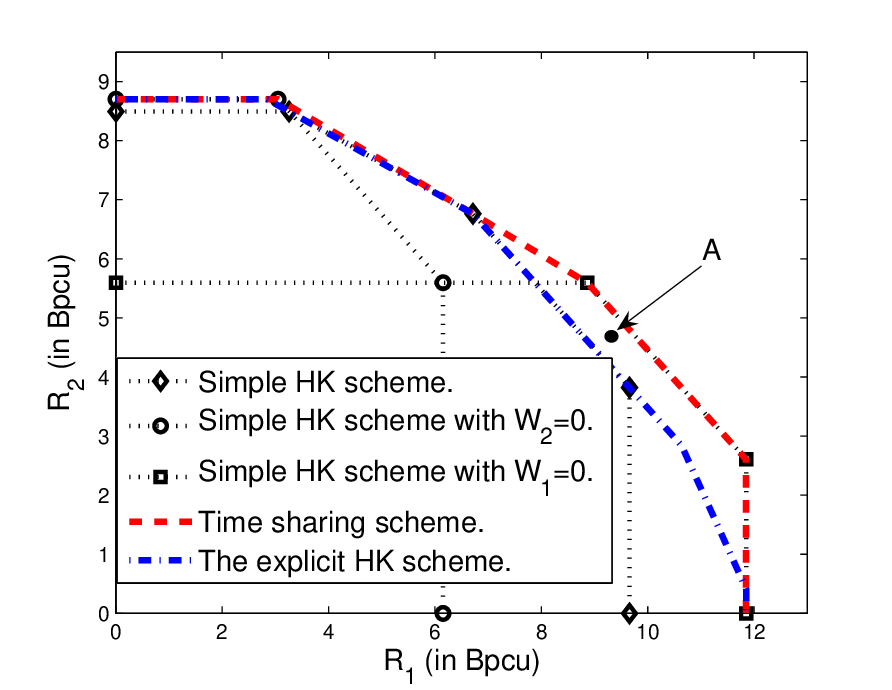}}
  \end{center}
\caption{A comparison of achievable rate regions for the simple and explicit HK schemes for the channel of Example~\ref{ex:achievable-region-of-simple-HK-scheme}.}
\label{fig:difference_with_timesharing}
\end{figure*}
\begin{thm}
\label{thm_approximate_capacity2}
The explicit HK coding scheme $\widetilde{\mathcal{HK}}$ has an achievable rate region that is within $(n_1^*, n_2^*)$ bits to the capacity region of the Gaussian MIMO IC, where $n_i^*$ is given by \eqref{eq_def_nis}.
\end{thm}

\begin{IEEEproof}
Applying Lemma 2 of \cite{CMG} to the Gaussian MIMO IC for the distribution $P_s$ of Definition \ref{def:coding-scheme-input-distribution}, it is easily seen that $ \mathcal{R}_{\textrm{HK}}^{G_c}(P_s) \subseteq \mathcal{R}_{\textrm{HK}}^{G_e}(P_s) \cup \mathcal{R}_{\textrm{HK}}^{G_e}(P_{s_1}) \cup \mathcal{R}_{\textrm{HK}}^{G_e}(P_{s_2})$. Hence, $ \mathcal{R}_{\textrm{HK}}^{G_c}(P_s) $ is an inner bound to the achievable rate region of the $\widetilde{\mathcal{HK}}$. Moreover, using the proof in Appendix \ref{pf:lem:rate-region-within-ni-bits} of Lemma \ref{lem:rate-region-within-ni-bits} that $\mathcal{R}_a(\mathcal{H},\bar{\rho})$ is an inner bound for $ \mathcal{R}_{\textrm{HK}}^{G_e}(P_s)$, we see that the corresponding inner bound for $ \mathcal{R}_{\textrm{HK}}^{G_c}(P_s) $ is the rate region defined by the inequalities in \eqref{eq:intermediate-rate-region-a} and \eqref{eq:intermediate-rate-region-c} (but not \eqref{eq:intermediate-rate-region-b} and \eqref{eq:intermediate-rate-region-d}) together with the \eqref{eq:rate-region-within-ni-bits3}-\eqref{eq:rate-region-within-ni-bits7}. Now, using the argument in the proof of Theorem \ref{thm_approximate_capacity}, it is clear that the achievable rate region of the explicit $\widetilde{\mathcal{HK}}$ scheme 
is within the smaller gap of $(n_1^\ast, n_2^\ast)$ of the outer bound $\mathcal{R}^u(\mathcal{H},\bar{\rho})$ of Lemma \ref{lem_upper_bound}, and hence it is within a gap of $(n_1^\ast, n_2^\ast)$ of the capacity region\footnote{In this explicit scheme the sub-rates for the private and public messages of the two transmitters are chosen from $\mathcal{R}_{\textrm{HK}}^{G_o}(P)$, with $P\in \{P_s,P_{s_1},P_{s_2}\}$ depending on the rate pair to be achieved, and where $\mathcal{R}_{\textrm{HK}}^{G_o}(P)$ can be computed from \eqref{eqhk} and \eqref{eq:DMC-sub-rate-region} by using the distribution $P$ in place of $P^*$.}. Note that for the SISO Gaussian IC, the explicit scheme is within one bit of the capacity region.
\end{IEEEproof}

We give an intuitive explanation of why the simple HK scheme $\mathcal{HK}^{(s)}$ cannot achieve the smaller gap. Suppose there exists a rate pair $(R_1^{'},R_2^{'})$ that satisfies all the constraints of $ \mathcal{R}_{\textrm{HK}}^{G_e}(P_s) $ except
\eqref{eq:compact-gaussian-rate-region-1b}, i.e.,
\begin{IEEEeqnarray}{rl}
I(X_1^g ; Y_1^g|W_1^g,W_2^g)+I(W_1^g;Y_2^g|X_2^g) <& R_1^{'} \nonumber\\
    \label{eq:when-H12-is-weak}
    \leq & I(X_1^g ; Y_1^g|W_2^g).~~~~~
\end{IEEEeqnarray}
The maximum value of $R_1\in \mathcal{R}_{\textrm{HK}}^{G_e}(P_s)$ in such a scenario is restricted only by the bound corresponding to \eqref{eq:compact-gaussian-rate-region-1a}. However, comparing the two sides of \eqref{eq:when-H12-is-weak} we see that, the first term on the left hand side of \eqref{eq:when-H12-is-weak} differs from that on the right hand side only due to the extra $W_1^g$ in the conditioning. If all of the power is allocated to the private message only (i.e., $X_1^g=U_1^g$ and $W_1^g=\phi$) then the first term on the left hand side alone is equal to the right hand side and \eqref{eq:when-H12-is-weak} cannot be true. So, some fraction of the total power available at $Tx_1$ is used to send $W_1^g$ which decreases the term $I(X_1^g ;Y_1^g|W_1^g,W_2^g)$. However, this decrease is more than the corresponding increase in the second term on the left hand side, $I(W_1^g;Y_2^g|X_2^g)$. Now, since $I(X_1^g ;Y_1^g|W_1^g,W_2^g)$ represents the information carried by the private message of the first user on the direct link and $I(W_1^g;Y_2^g|X_2^g)$ represents the information carried by only the public message of user 1 on the cross link in the absence of $X_2$, this in turn suggests that the cross link from $Tx_1$ to $Rx_2$ is {\it weaker} in some sense than the direct link.

Clearly, when \eqref{eq:when-H12-is-weak} is true the rate pair $(R_1^{'},R_2^{'})$ is not achievable by the simple HK scheme. The main drawback  of the encoding technique in the above scenario is therefore that a significant fraction of the power is spent to send some common information ($W_1^g$) through a weak channel to a receiver $(Rx_2)$ where the message is not even desirable. Intuitively, it appears that instead of wasting power on a weak channel, $Tx_1$ should not send any public information at all, i.e., it must set $K_{1w}=\mathbf{0}$ and assign all of its power to the private message. As mentioned earlier, if we put $X_1^g=U_1^g$ and $W_1^g=0$ in \eqref{eq:when-H12-is-weak}, the strict inequality becomes an equality, i.e.,
\begin{IEEEeqnarray*}{rl}
    I(X_1^g ; Y_1^g|W_1^g, W_2^g)+I(W_1^g;Y_2^g|X_2^g)= &~I(U_1^g ; Y_1^g|W_2^g)\\
     = & ~I(X_1^g;Y_1^g|W_2^g),
\end{IEEEeqnarray*}
and the two bounds \eqref{eq:compact-gaussian-rate-region-1a} and \eqref{eq:compact-gaussian-rate-region-1b} become identical. With such a covariance split it turns out that the rate pair $(R_1^{'},R_2^{'})$ is indeed achievable. 


\begin{rem}
It is worth pointing out the differences between the explicit coding scheme of this paper specialized to the SISO IC and that in \cite{ETW1}, where a simple coding scheme was also suggested to characterize the capacity region of the SISO IC within one bit. The authors in \cite{ETW1} use a linear superposition coding scheme where each users private and public messages are encoded using independent Gaussian random codewords with powers $P_{iu}$ and $(P_i-P_{iu})$, respectively for $i=1,2$. Here, $P_i$ is the total average power of $Tx_i$ and $P_{iu}$ depends on the cross channel coefficients as follows (see (57) and (58) of \cite{ETW1})
\begin{equation}\label{eq:ETW-power-split-scheme}
    P_{iu}=\min \{P_i, \frac{1}{\|H_{ij}\|^2}\}, ~i\neq j\in \{1,2\}.
\end{equation}
In the notation of the present paper this coding scheme is $\mathcal{HK}\left(\left\{\frac{P_{1u}}{P_1},\left(1-\frac{P_{1u}}{P_1}\right),\frac{P_{2u}}{P_2},
\left(1-\frac{P_{2u}}{P_2}\right)\right\}\right)$, when $P_i\geq \frac{1}{\|H_{ij}\|^2}$. On the other hand, it is $\mathcal{HK}\left(\left\{1,0,\frac{P_{2u}}{P_2},\left(1-\frac{P_{2u}}{P_2}\right)\right\}\right)$, when only $P_1 < \frac{1}{\|H_{12}\|^2}$, $\mathcal{HK}\left(\left\{\frac{P_{1u}}{P_1},\left(1-\frac{P_{1u}}{P_1}\right),1,0\right\}\right)$, when only $P_2< \frac{1}{\|H_{21}\|^2}$ and it is $\mathcal{HK}(\{1,0,1,0\})$, when $P_i\leq \frac{1}{\|H_{ij}\|^2}$ for both $i=1,2$. So, depending on the channel coefficients the coding scheme is equivalent to one of the four schemes just described. However, for a given channel the coding scheme and power allocation of \cite{ETW1} is fixed and does not change with the rate pair to be achieved. In contrast, the explicit coding scheme of this paper utilizes one of the three different power splitting schemes depending on the rate pair to be achieved while also achieving a constant gap to capacity of one bit.

\end{rem}

\begin{figure*}[htp]
  \begin{center}
    \subfigure[Reversing the direction of information flow in the $(M_1,N_1,M_2,N_2)$ IC.]{\label{fig_reciprocity-a}\includegraphics[scale=.25]{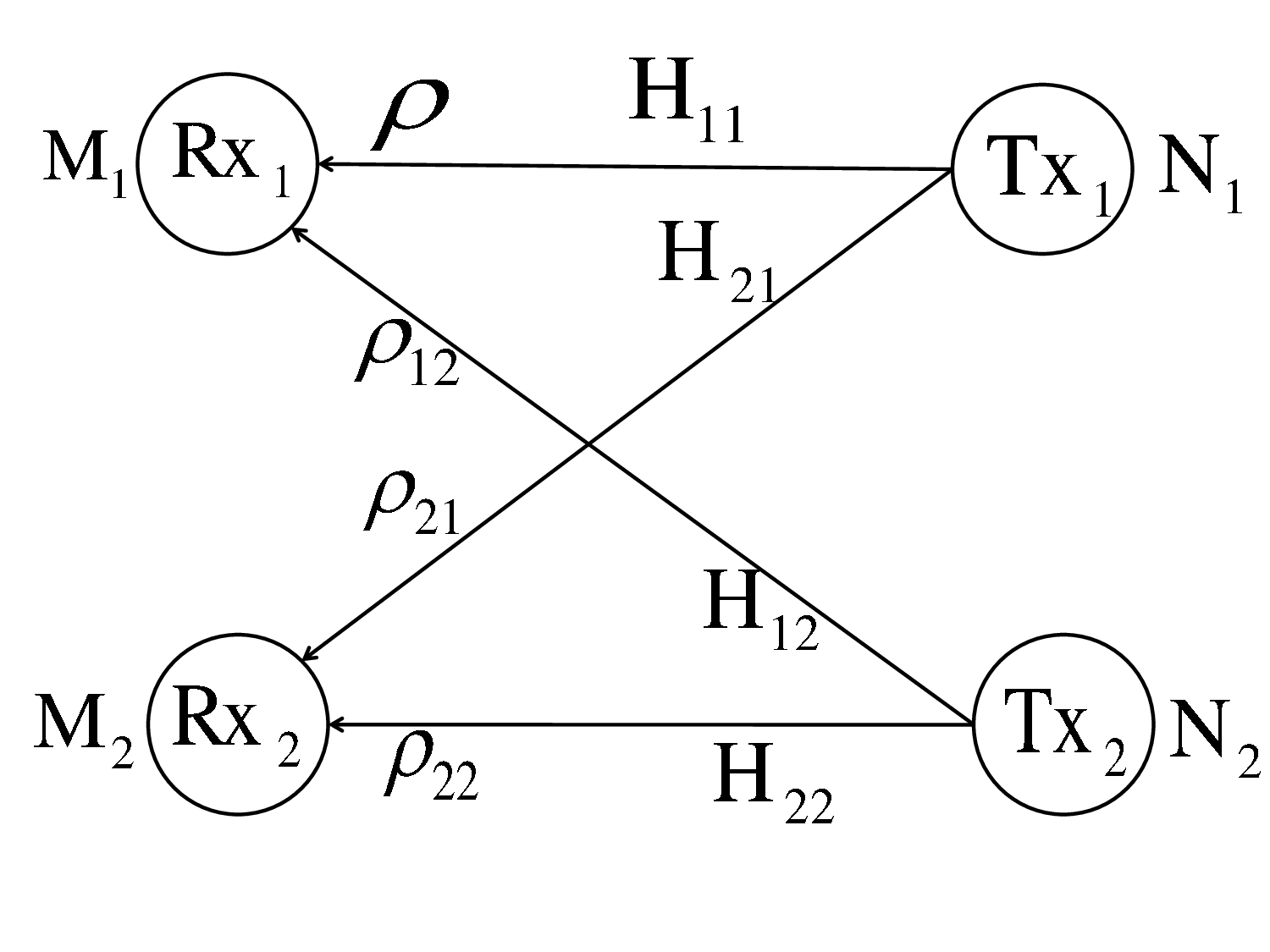}}
    \subfigure[Equivalent channel with information flowing in the forward direction.]{\label{fig_reciprocity-b}\includegraphics[scale=0.25]{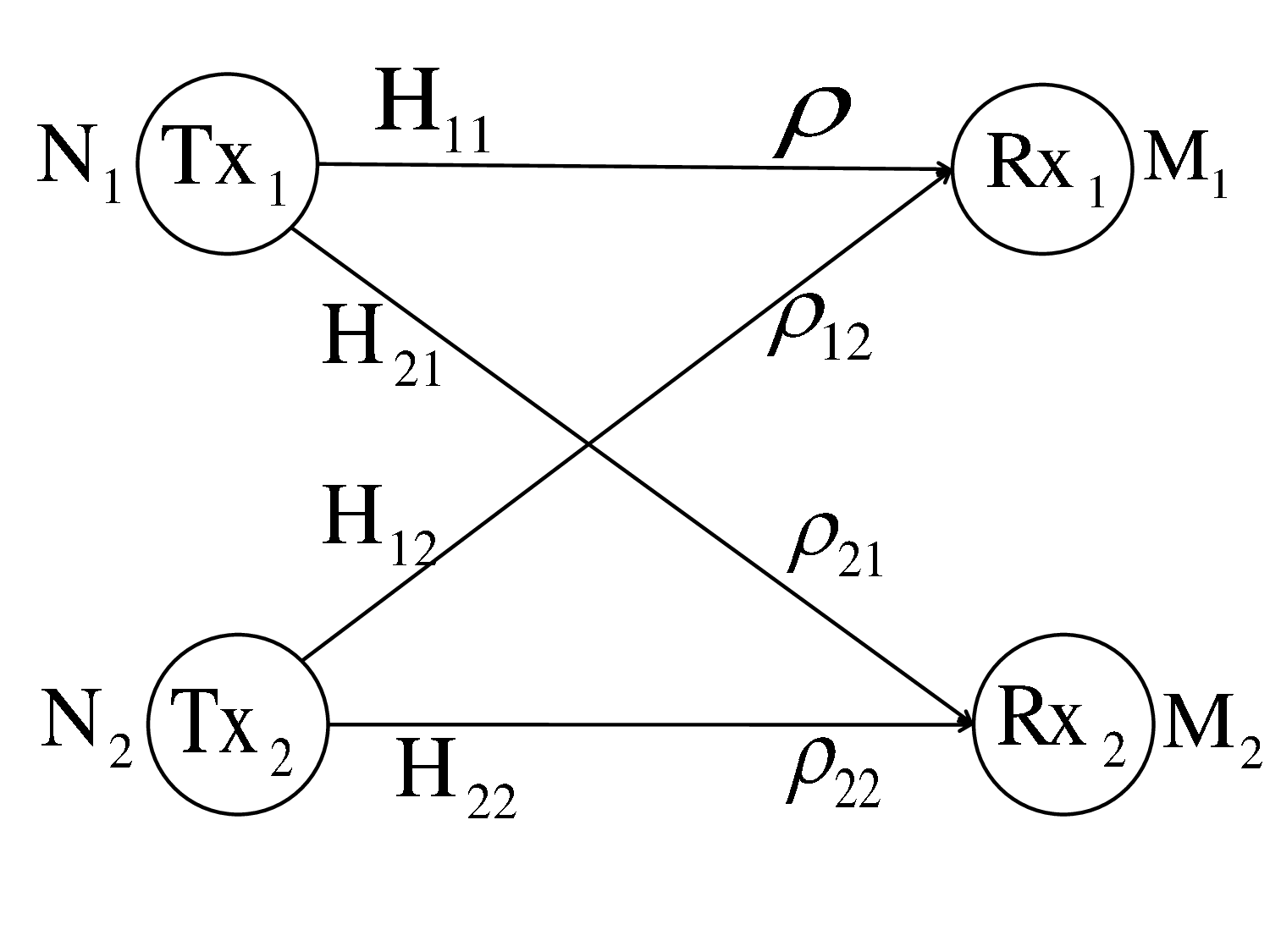}}
  \end{center}
\caption{Information flowing in the reverse direction in a two-user MIMO IC and its corresponding forward information flow model.}
\label{fig_reciprocity}
\end{figure*}

\begin{ex}
\label{rem:time-sharing-difference}
Fig. \ref{fig:difference_with_timesharing-a} depicts the achievable rate regions of the three simple HK schemes $\mathcal{HK}^{(s)}, \mathcal{HK}^{(s_1)}$, and $\mathcal{HK}^{(s_2)}$ for the channel of Example \ref{ex:achievable-region-of-simple-HK-scheme}. The rate region achievable by the explicit scheme that chooses one of $\mathcal{HK}^{(s)}, \mathcal{HK}^{(s_1)}$, and $\mathcal{HK}^{(s_2)}$ is the union of these three regions.
In Fig. \ref{fig:difference_with_timesharing-b} the rate region bounded by the dotted-dashed line represents the compact rate region $\mathcal{R}_{\textrm{HK}}^{G_c}(P_s) $ that is contained in the achievable rate region of the explicit HK scheme. Moreover, it strictly contains $\mathcal{R}_{\textrm{HK}}^{G_e}(P_s) $. The dashed line represents the rate region, $\mathcal{R}_{TS}$, achievable by time sharing between the three constituent schemes $\mathcal{HK}^{(s)}, \mathcal{HK}^{(s_1)}$, and $\mathcal{HK}^{(s_2)}$. While the compact rate region $\mathcal{R}_{\textrm{HK}}^{G_c}(P_s) $ is within $(n_1^\ast, n_1^\ast)$ bits of the capacity region, it is not clear if using the three distributions and time sharing leads to improved performance in terms of guaranteeing a smaller gap to the capacity region, i.e., if the gap between $\mathcal{R}_{TS}$ and the capacity region can be shown to be smaller.

\end{ex}

\begin{rem}
Capacity of the SIMO IC within one bit: On a $(1,N_1,1,N_2)$ IC, $n_1^*=n_2^*=1$, thus the explicit $\widetilde{\mathcal{HK}}$ scheme can achieve a rate region which is within 1 bit of the capacity region for any SNRs, INRs and the channel vectors. This result is different from that reported in~\cite{Sriram_Jafar} where the exact sum capacity of the {\em strong} SIMO IC with $\|H_{ii}\|^2\leq \|H_{ij}\|^2$ for $1\leq i\neq j\leq 2$, was characterized. While \cite{Sriram_Jafar} provides the exact sum capacity for the strong SIMO IC, our 1 bit approximation is valid for all channel coefficients. Further, this approximation is tighter than that reported in~\cite{TT} and \cite{Wang_Tse}, where the capacity approximation within $(N_1, N_2)$ bits was proved.
\end{rem}


\begin{rem}
The achievable rate region in \cite{TT} was proved to be within $(I(X_2 ; S_2|\tilde{U}_2), I(X_1 ; S_1|\tilde{U}_1))$ of the capacity region, which for the MIMO Gaussian IC was shown to be within $(N_1,N_2)$ bits of the capacity region. However, in this latter channel, the gap in terms of conditional mutual informations obtained in \cite{TT} can in fact be easily shown to be within $(m_{21}, m_{12})$ bits of the capacity region, or within $(\min \{M_2,N_1\}, \min \{M_1,N_2\}) $ in the case of full rank cross channel matrices. The $(n_1^*, n_2^*)$ gap proved in this paper for the explicit HK scheme is identical to the smaller gap of \cite{TT}, but for others, it is larger. However, as mentioned previously, the achievable scheme in \cite{TT} requires the consideration of all possible input distributions and all possible time sharing schemes (without shedding light on what simple or explicit scheme, if any, among those are good) whereas the explicit HK scheme here requires just three distributions, no time-sharing and is approximate-capacity-optimal.
\end{rem}

\begin{rem}
Moreover, the explicit inner and outer bounds to the capacity region permit further analysis at high SNR because they are within a constant gap of each other.  For instance, the so called generalized degrees of freedom region was obtained by the authors in the companion paper \cite{Karmakar-Varanasi-GDoF}, generalizing the result for the SISO IC in \cite{ETW1}, and providing further insights about the explicit (or simple) HK scheme at high SNR. Furthermore, for the same reason, in the quasi-static fading MIMO IC, it is possible to even obtain the rate-reliability tradeoff at high SNR in the form of the fundamental diversity-multiplexing tradeoff (DMT), and this too was obtained by the authors for the Gaussian MIMO Z and MIMO IC channels in \cite{Sanjay_Varanasi_ZIC_DMT} and \cite{Sanjay_Varanasi_IC_DMT}, respectively.
\end{rem}


\begin{rem}
\label{rem:compound-ic-proof}
Recall the rate region $\mathcal{R}_{\textrm{in}}(P_1^*)=\mathcal{R}_{\textrm{CMG}}(P_1^*)$  and the discussion in Remark \ref{rem:compound-ic-result}. The distinguishing aspect of \cite{raja_prabhakaran_viswanath} 
from that in \cite{CMG} is that the rate region $\mathcal{R}_{\textrm{in}}(P_1^*)$ for each $P_1^*$ was shown in \cite{raja_prabhakaran_viswanath} to be achievable by a single input distribution. Now, letting $P_1^*$ be the joint distribution of $X_1, X_2, W_1, W_2 $ obtained from $P_s$ and denoting the resulting rate region as $\mathcal{R}_{\textrm{CMG}}(P_s^1)$, we have since $\mathcal{R}_{\textrm{HK}}^{G_e}(P_s) \subseteq \mathcal{R}_{\textrm{CMG}}(P_s^1) $ \cite{CMG}, that $\mathcal{R}_{\textrm{CMG}}(P_s^1)$ (and hence the scheme of \cite{raja_prabhakaran_viswanath}) is within a gap of $(n_1, n_2)$ of the capacity region. However, since $\mathcal{R}_{\textrm{CMG}}(P_s^1) \subseteq \mathcal{R}_{\textrm{HK}}^{G_c}(P_s) $, it is not clear that a gap of $(n_1^\ast, n_2^\ast)$ to the capacity region is achievable by the scheme of \cite{raja_prabhakaran_viswanath} or by some other single distribution.

\end{rem}

\subsection{Reciprocity of the approximate capacity region}
\label{subsection_reciprocity}
For a MIMO point-to-point channel with possibly an unequal number of antennas at the source and destination nodes, the capacity remains unchanged when the information flows in the opposite direction (i.e., the roles of the transmitter and the receiver are interchanged) and this property of the channel is called reciprocity (see Remark 1 of \cite{Telatar}). In a network such as the MIMO IC, it is of interest to know if the reciprocity property holds in the sense that some approximation of capacity (since the capacity is not known) remains unchanged if the information flows in the opposite direction so that the roles of the transmitter and the receiver of each of the two transmit-receive pairs are interchanged. 
For instance, 
the so-called degrees of freedom (DoF) region of a $(M_1,N_1,M_2,N_2)$ MIMO IC found in \cite{JFak} is the same as that of a $(N_1,M_1,N_2,M_2)$ IC. In this section, we prove a reciprocity result for the $(M_1,N_1,M_2,N_2)$ MIMO IC by showing that reciprocity holds in the much stronger constant-gap-to-capacity sense.

Fig. \ref{fig_reciprocity-a} illustrates an $(M_1,N_1,M_2,N_2)$ MIMO IC with channel parameters $H$ and $\bar{\rho}$ with the roles of the transmitters and receivers interchanged so that information flows in the reverse direction. Fig. \ref{fig_reciprocity-b} shows its equivalent model where the information flows in the forward direction. Clearly, the capacity of the reverse channel is the same as that of $\mathcal{IC}\left(\mathcal{H}^r,\bar{\rho}^r\right)$ where $\mathcal{H}^r=\{H_{11}^T, H_{21}^T, H_{12}^T, H_{22}^T\}$ and $\bar{\rho}^r=  [\rho,\rho_{21},\rho_{12},\rho_{22}]$. The capacity region of the reverse channel is denoted as $\mathcal{C}\left(\mathcal{H}^r,\bar{\rho}^r\right)$.

Let us define the counterparts in the reverse channel of the capacity gap parameters of the forward channel in \eqref{eq_def_nis} as
\begin{equation*}
\label{eq_def_mis}
m_i^* \triangleq \min\{M_i,N_s\}\log(N_x)+\tilde{m}_{ij},  \qquad 1\leq i\neq j\leq 2,
\end{equation*}
where  $N_s = (N_1+N_2)  $, $ N_x = \max\{N_1,N_2\}$ and $\tilde{m}_{ij}=m_{ij}\log\left(\frac{N_j+1}{N_j}\right)$ for $1\leq i\neq j\leq 2.$

To prove the reciprocity in the constant-gap-to-capacity sense, the capacity regions $\mathcal{C}\left(\mathcal{H}, \bar{\rho}\right)$ and $\mathcal{C}\left(\mathcal{H}^r,\bar{\rho}^r\right)$ must be shown to be within a constant gap of each other. The key to proving this result is the following lemma.
\begin{lemma}
\label{lem_reciprocity_ub}
The outer bound $ \mathcal{R}^u\left(\mathcal{H}, \bar{\rho}\right) $ from Lemma \ref{lem_upper_bound} of the forward channel $\mathcal{IC}\left(\mathcal{H}, \bar{\rho}\right)$ and the outer bound $ \mathcal{R}^u\left(\mathcal{H}^r, \bar{\rho}^r\right) $ (obtained in the same way as in Lemma \ref{lem_upper_bound} but for the reverse channel $\mathcal{IC}\left(\mathcal{H}^r,\bar{\rho}^r\right)$) 
define the same set of rate pairs, i.e.,
\begin{equation*}
\label{eq_reciprocity_ub}
\mathcal{R}^u\left(\mathcal{H}, \bar{\rho}\right)=\mathcal{R}^u\left(\mathcal{H}^r, \bar{\rho}^r\right).
\end{equation*}
\end{lemma}

\begin{IEEEproof}
The proof is given in Appendix~\ref{pf_lem_reciprocity_ub}.
\end{IEEEproof}

Theorem \ref{thm_approximate_capacity2} proves that the explicit HK scheme, $\widetilde{\mathcal{HK}}$, achieves a rate region on $\mathcal{IC}(\mathcal{H},\bar{\rho})$ which is within $(n_1^*, n_2^*)$ bits of $\mathcal{R}^u(\mathcal{H},\bar{\rho})$, which in turn contains its capacity region. Clearly, the counterpart of this explicit HK coding scheme for the reverse channel (with suitable changes in the channel matrices, INRs and the number of antennas) can achieve a rate region on $\mathcal{IC}\left(\mathcal{H}^r,\bar{\rho}^r\right)$ which is within $(m_1^*, m_2^*)$ bits of
$\mathcal{R}^u\left(\mathcal{H}^r, \bar{\rho}^r\right)$. However, since
$ \mathcal{R}^u\left(\mathcal{H}, \bar{\rho}\right)=\mathcal{R}^u\left(\mathcal{H}^r, \bar{\rho}^r\right) $ from Lemma~\ref{lem_reciprocity_ub}, we have that the capacity regions $\mathcal{C}\left(\mathcal{H}, \bar{\rho}\right)$ and $\mathcal{C}\left(\mathcal{H}^r,\bar{\rho}^r\right)$ within a constant gap as shown in  the following theorem.
\begin{thm}
\label{thm_reciprocity}
The capacity regions of $\mathcal{IC}(\mathcal{H},\bar{\rho})$ and $\mathcal{IC}(H^r,\bar{\rho}^r)$ are within $(\max\{m_1^*,n_2^*\}, \max\{m_2^*,n_2^*\})$ bits of each other, i.e., if $(R_1,R_2)\in \mathcal{C}(\mathcal{H},\bar{\rho})$, then there exists a rate pair $(R_1^r,R_2^r)\in \mathcal{C}(\mathcal{H}^r,\bar{\rho}^r)$ such that $ |(R_i-R_i^r)|\leq \max\{m_i^*,n_i^*\}, \; i \in \{1,2\} $.
\end{thm}
\begin{IEEEproof}
Let $(R_1,R_2)\in \mathcal{C}(\mathcal{H},\bar{\rho})$. From Theorem~\ref{thm_approximate_capacity2}, there exist a rate pair $(\hat{R}_1,\hat{R}_2)\in \mathcal{R}^u(\mathcal{H},\bar{\rho})$ such that
\begin{equation}
\label{eq_pf_reciprocity_1}
0\leq \hat{R}_i-R_i \leq n_i^*, \; i \in \{1,2\}.
\end{equation}
Since from Lemma~\ref{lem_reciprocity_ub} $(\hat{R}_1,\hat{R}_2)\in \mathcal{R}^u(\mathcal{H}^r,\bar{\rho}^r)$, applying Theorem~\ref{thm_approximate_capacity2} to $\mathcal{IC}(\mathcal{H}^r,\bar{\rho}^r)$, there exists a rate pair $(R_1^r,R_2^r)\in \mathcal{C}(\mathcal{H}^r,\bar{\rho}^r)$ such that
\begin{equation}
\label{eq_pf_reciprocity_2}
0\leq \hat{R}_i-R_i^r\leq m_i^*, \; i \in \{1,2\}.
\end{equation}
Note that \eqref{eq_pf_reciprocity_1} and \eqref{eq_pf_reciprocity_2} provide ranges of $R_i$ and $R_i^r$ and the magnitude of the difference between them is maximum when one takes its largest value and the other its smallest, i.e.,
\begin{IEEEeqnarray*}{rl}
|R_i-R_i^r|_{max}=m_i^*~\textrm{or}~n_i^*,
\end{IEEEeqnarray*}
which proves the theorem.
\end{IEEEproof}



\section{Conclusion}
\label{sec_conclusion}
An approximate capacity region of the two-user MIMO IC with an arbitrary number of antennas at each node is characterized by obtaining explicit inner and outer bounds that are within a constant gap of each other. It is shown that a simple and an explicit HK coding schemes that can be seen to inherently perform a form of joint interference alignment in the signal space and in the signal level (see Section II.C of the companion paper \cite{Karmakar-Varanasi-GDoF} for this interpretation) can achieve the capacity region of the MIMO IC to within the constant gap. For a class of MIMO ICs, this gap is the tightest approximation to the capacity region of the MIMO IC found to date and this includes the SIMO ICs for which the gap is one bit independently of the numbers of antennas at the receivers. The explicit upper and lower bounds to the capacity region are used to prove the reciprocity of the MIMO IC in the constant-gap-to-capacity sense.

\appendices

\section{A matrix inequality}
\label{app:lem_conditional_entropy}
\begin{lemma}
\label{lem_partial_order_sequence}
Let $0\preceq G_1\preceq G_2$ and $0\preceq A$ are p.s.d. matrices of size $n$, then for any given $\pi\in \mathbb{R}^+$
\begin{equation*}
    G_1\left(I+\pi G_1AG_1\right)^{-1}G_1 \preceq  G_2\left(I+\pi G_2AG_2\right)^{-1}G_2.
\end{equation*}
\end{lemma}
\begin{IEEEproof}
Let $\epsilon\in \mathbb{R}^+$, $G_{1\epsilon}=\left(G_1+\epsilon I \right)$ and $G_{2\epsilon}=\left(G_2+ \epsilon I \right)$. For any such $\epsilon$, we have
\begin{IEEEeqnarray*}{rl}
    G_{2\epsilon}\succeq & G_{1\epsilon}\succ 0, \\ \textrm{or},~
    G_{1\epsilon}^{-2}\succeq & G_{2\epsilon}^{-2}\succ 0, \\  \textrm{or},~
    \left( G_{1\epsilon}^{-2}+\pi A\right)\succeq &\left(G_{2\epsilon}^{-2}+\pi A\right)\succ 0, \\ \textrm{or},~
    \left( G_{1\epsilon}^{-2}+\pi A\right)^{-1}\preceq &\left(G_{2\epsilon}^{-2}+\pi A\right)^{-1},
\end{IEEEeqnarray*}
which in turn imply that
\begin{equation*}
     G_{1\epsilon}\left(I+\pi G_{1\epsilon}AG_{1\epsilon}\right)^{-1}G_{1\epsilon} \preceq  G_{2\epsilon}\left(I+\pi G_{2\epsilon}AG_{2\epsilon}\right)^{-1}G_{2\epsilon}.
\end{equation*}

From the definition of partial order between p.s.d. matrices we get, $\forall ~x\in \mathbb{C}^{1\times n}$,
\begin{IEEEeqnarray*}{l}
x\left(G_{1\epsilon}\left(I+\pi G_{1\epsilon}AG_{1\epsilon}\right)^{-1}G_{1\epsilon}\right)x^\dagger \\
~~~~~~~~~~~~~~~~
\leq \  x\left(G_{2\epsilon}\left(I+\pi G_{2\epsilon}AG_{2\epsilon}\right)^{-1}G_{2\epsilon}\right)x^\dagger;\\
{\Rightarrow}~\lim_{\epsilon \to 0}x\left(G_{1\epsilon}\left(I+\pi G_{1\epsilon}AG_{1\epsilon}\right)^{-1}G_{1\epsilon}\right)x^\dagger \\
~~~~~~~~~~~~~~~~\leq \
\lim_{\epsilon \to 0}~x\left(G_{2\epsilon}\left(I+\pi G_{2\epsilon}AG_{2\epsilon}\right)^{-1}G_{2\epsilon}\right)x^\dagger,
\end{IEEEeqnarray*}
 where the last step follows from the fact that for any $G_i$, $A$ and $\pi$ as defined above and for any sequence of positive real numbers $\{\epsilon_n\}_{n=1}^\infty$ with $\epsilon \to 0$ as $n\to \infty$ we have
\begin{IEEEeqnarray*}{l}
\lim_{n\to \infty}\left(G_{i\epsilon_n}\left(I+\pi G_{i\epsilon_n}AG_{i\epsilon_n}\right)^{-1}G_{i\epsilon_n}\right)\\
~~~~~~~~~~~~~~~=~\left(G_i\left(I+\pi G_iAG_i\right)^{-1}G_i\right).
\end{IEEEeqnarray*}
Substituting their limits in the last equation we get
\begin{IEEEeqnarray*}{l}
x\left(G_1\left(I+\pi G_1AG_1\right)^{-1}G_1\right)x^\dagger \\
~~~~~~~~~~~~~~~\leq \  x\left(G_2\left(I+\pi G_2AG_2\right)^{-1}G_2\right)x^\dagger,
\end{IEEEeqnarray*}
for all $x\in \mathbb{C}^{1\times n}$. Invoking the definition of partial ordering once again, the lemma is proved.
\end{IEEEproof}



\section{Proof of Lemma~\ref{lem:rate-region-within-ni-bits}}
\label{pf:lem:rate-region-within-ni-bits}
The idea is to show that the bounds in \eqref{eq:rate-region-within-ni-bits1}-\eqref{eq:rate-region-within-ni-bits7} are obtained by replacing the right hand sides of \eqref{eq:compact-gaussian-rate-region-1a}-\eqref{eq:compact-gaussian-rate-region-1i} by their respective lower bounds, namely, the right hand sides of the bounds describing $\mathcal{R}_a(\mathcal{H},\bar{\rho})$).
To this end, we derive some common inequalities which will be used throughout the proof.

First, since $ K_{iu} = \left(I_{M_i}+\rho_{ij} H_{ij}^{\dagger}H_{ij}\right)^{-1} \succ \mathbf{0} $, we have from the definition of p.s.d. matrices~\cite{HJ} that
\begin{IEEEeqnarray}{rl}
\left(\rho_{ij}H_{ij}K_{iu}H_{ij}^{\dagger}\right)= &\left(\frac{\rho_{ij}}{M_{i}}H_{ij}\left(I_{M_i}+\rho_{ij} H_{ij}^{\dagger}H_{ij}\right)^{-1}H_{ij}^{\dagger}\right)\nonumber\\
\label{eq_pf_achieve_lemma_t1}
& \succeq \mathbf{0}.
\end{IEEEeqnarray}
Next, with  $H_{ij}=U_{ij}\Sigma_{ij}V_{ij}^\dagger$ being the singular value decomposition of $H_{ij}$ \cite{HJ}, we have for any non-zero $x\in \mathbb{C}^{1\times N_j}$ that
\begin{IEEEeqnarray*}{l}
x\left(\frac{\rho_{ij}}{M_{i}}H_{ij}\left(I_{M_i}+\rho_{ij} H_{ij}^{\dagger}H_{ij}\right)^{-1}H_{ij}^{\dagger}\right)x^\dagger\\
=\frac{\rho_{ij}}{M_{i}}\left(xU_{ij}\right)\Sigma_{ij}\left(I_{M_i}+\rho_{ij} \Sigma_{ij}^{\dagger}\Sigma_{ij}\right)^{-1}\Sigma_{ij}^\dagger\left(x U_{ij}\right)^{\dagger},\\
\stackrel{(a)}{\leq}  \frac{1}{M_i}(xU_{ij})(xU_{ij})^\dagger \leq \frac{xx^\dagger}{M_i},
\end{IEEEeqnarray*}
where step $(a)$ follows from the fact that $\rho_{ij}\Sigma_{ij}\left(I_{M_i}+\rho_{ij} \Sigma_{ij}^{\dagger}\Sigma_{ij}\right)^{-1}\Sigma_{ij}^\dagger \preceq I_{N_j}$. Hence, we have $\rho_{ij}H_{ij}K_{iu}H_{ij}^\dagger {\preceq} \frac{1}{M_i}I_{N_j} $, so that
\begin{IEEEeqnarray}{rl}
\label{eq_pf_achieve_lemma_t2}
\log\det \left(I_{N_j}+\rho_{ij}H_{ij}K_{iu}H_{ij}^{\dagger}\right)\leq & m_{ij}\log\left(\frac{1+M_i}{M_i}\right) \\
=& \hat{m}_{ij},~\textrm{for}~ 1\leq i\neq j\leq 2.  \nonumber
\end{IEEEeqnarray}

As a first step towards deriving the lower bounds, in what follows, we shall first derive lower bounds for the different mutual information terms of \eqref{eq:Is-for-Gaussian-input-1}-\eqref{eq:Is-for-Gaussian-input-7}. For the bound in \eqref{eq:Is-for-Gaussian-input-1}, we obtain
\begin{IEEEeqnarray}{l}
I(X_1^g;Y_1^g|W_1^g,W_2^g)\nonumber\\
=\log \det\left(\frac{\rho_{11}}{M_1}H_{11}K_1H_{11}^{\dagger}+\frac{\rho_{21}}{M_2}H_{21}K_2H_{21}^{\dagger}+I_{N_1}\right)\nonumber\\
-\log \det \left(\rho_{21}H_{21}K_{2u}H_{21}^{\dagger}+I_{N_1}\right),\nonumber\\
\stackrel{(a)}{\geq} \log \det\left(\frac{\rho_{11}}{M_1}H_{11}K_1 H_{11}^{\dagger}+\frac{1}{M_1}I_{N_1}\right)-\hat{m}_{21},\nonumber \\
=\log \det\left(\rho_{11}H_{11}K_1H_{11}^{\dagger}+I_{N_1}\right)\nonumber\\
\label{eq_pf_lem_achievable_a1}
~~~~~~~~~~~~~~~~~~~~~-(m_{11}\log(M_1)+\hat{m}_{21}),
\end{IEEEeqnarray}
where step $(a)$ follows from \eqref{eq_pf_achieve_lemma_t1} and \eqref{eq_pf_achieve_lemma_t2} and the fact that $\log \det (.)$ is a monotonically increasing function over the cone of positive-definite matrices with respect to partial ordering. Similarly,
\begin{IEEEeqnarray}{rl}
I(X_2^g;Y_2^g|W_2^g,W_1^g)~\geq & \log \det\left(\rho_{22}H_{22}K_2H_{22}^{\dagger}+I_{N_2}\right)\nonumber\\
\label{eq_pf_lem_achievable_a2}
&~~~~~ -(m_{22}\log(M_2)+\hat{m}_{12}).
\end{IEEEeqnarray}

For the bounds in \eqref{eq:Is-for-Gaussian-input-3} and \eqref{eq:Is-for-Gaussian-input-4}, using similar reasoning, we obtain
\begin{IEEEeqnarray}{rl}
I(W_2^g;Y_1^g|X_1^g) {\geq}& \log \det\left(\rho_{21}H_{21}H_{21}^{\dagger}+I_{N_1}\right)\nonumber\\
&~~~~-(m_{21}\log(M_2)+\hat{m}_{21}),
\label{eq_pf_lem_achievable_c1} \\
I(W_1^g;Y_2^g|X_2^g)\geq &\log \det \left(\rho_{12} H_{12}H_{12}^{\dagger}+ I_{N_2}\right)\nonumber\\
&~~~~-(m_{12}\log(M_1)+\hat{m}_{12}),
\label{eq_pf_lem_achievable_c2} \end{IEEEeqnarray}
\begin{IEEEeqnarray}{rl}
I(X_1^g;Y_1^g|W_2^g) \geq & \log \det\left(\rho_{11} H_{11}H_{11}^{\dagger}+I_{N_1}\right)\nonumber \\
&~~~~-(m_{11}\log(M_1)+\hat{m}_{21}),
\label{eq_pf_lem_achievable_d1} \\
I(X_2^g;Y_2^g|W_1^g) \geq & \log \det\left(\rho_{22} H_{22}H_{22}^{\dagger}+I_{N_2}\right)\nonumber\\
&~~~~-(m_{22}\log(M_2)+\hat{m}_{12}).
\label{eq_pf_lem_achievable_d2}
\end{IEEEeqnarray}

Next, for the bound in \eqref{eq:Is-for-Gaussian-input-7} we have
\begin{IEEEeqnarray}{l}
I(X_1^g,W_2^g;Y_1^g)\nonumber\\
= \log \det\left(\frac{\rho_{21}}{M_2}H_{21} H_{21}^{\dagger}+\frac{\rho_{11}}{M_1}H_{11}H_{11}^{\dagger}+I_{N_1}\right),\nonumber\\
~~~~~~~~~~~~-\log \det \left(\rho_{21}H_{21}K_{2u}H_{21}^{\dagger}+I_{N_1}\right),\nonumber\\
\stackrel{(a)}{\geq}  \log \det\left(\frac{\rho_{21}}{M_{x}}H_{21} H_{21}^{\dagger}+\frac{\rho_{11}}{M_{x}}H_{11}H_{11}^{\dagger}+\frac{1}{M_{x}}I_{N_1}\right) -\hat{m}_{21}, \nonumber \\
=\log \det\left(\rho_{21}H_{21} H_{21}^{\dagger}+\rho_{11}H_{11}H_{11}^{\dagger}+I_{N_1}\right)\nonumber\\
~~~~~~~~~~~~-(\min\{N_1,M_s\}\log(M_x)+\hat{m}_{21}),
\label{eq_pf_lem_achievable_g2}
\end{IEEEeqnarray}
where step $(a)$ follows from equation \eqref{eq_pf_achieve_lemma_t2} and the last step follows because the rank of the matrix $\left(\frac{\rho_{21}}{M_{x}}H_{21} H_{21}^{\dagger}+\frac{\rho_{11}}{M_{x}}H_{11}H_{11}^{\dagger}\right)$ cannot be larger than $\min\{N_1,M_s\}$, with $M_s$ and $M_x$ as defined in Section~\ref{sec_approximate_capacity}. Similarly, interchanging indices 1 and 2, we have
\begin{IEEEeqnarray}{rl}
I(X_2^g,W_1^g;Y_2^g)\geq &\log \det\left(\rho_{12} H_{12}H_{12}^{\dagger}+\rho_{22} H_{22}H_{22}^{\dagger}+I_{N_2}\right)\nonumber \\
\label{eq_pf_lem_achievable_g1}
&~~~-(\min\{N_2,M_s\}\log(M_x)+\hat{m}_{12}).
\end{IEEEeqnarray}

For the bound in \eqref{eq:Is-for-Gaussian-input-5}, using similar reasoning, we have
\begin{IEEEeqnarray}{l}
I(X_1^g,W_2^g;Y_1^g|W_1^g)\nonumber\\
 \geq  \log \det\left(\rho_{11}H_{11}K_1H_{11}^{\dagger}+\rho_{21}H_{21}H_{21}^{\dagger}+I_{N_1}\right)\nonumber\\
~~~~~-(\min\{N_1,M_s\}\log(M_x)+\hat{m}_{21}), \label{eq_pf_lem_achievable_e1} \\
I(X_2^g,W_1^g;Y_2^g|W_2^g) \nonumber\\
\geq  \log \det \left(I_{N_2}+\rho_{12} H_{12}H_{12}^{\dagger}+\rho_{22} H_{22}K_2H_{22}^{\dagger}\right)\nonumber\\
~~~~~- (\min\{N_2,M_s\}\log(M_x)+\hat{m}_{12}). \label{eq_pf_lem_achievable_e2}
\end{IEEEeqnarray}

Now, consider the two bounds on $R_1$ from \eqref{eq:compact-gaussian-rate-region-1a} and \eqref{eq:compact-gaussian-rate-region-1b} which we restate here for convenience.
\begin{IEEEeqnarray}{ll}
R_1 \leq & I(X_1^g;Y_1^g|W_2^g) \quad {\rm and} \label{rewritecgrr1a} \\
R_1 \leq & I(X_1^g;Y_1^g|W_1^g,W_2^g) + I(W_1^g;Y_2^g|X_2^g) . \label{rewritecgrr1b}
\end{IEEEeqnarray}
Substituting the lower bounds for the terms on the right hand sides of \eqref{rewritecgrr1a} and \eqref{rewritecgrr1b} obtained in
\eqref{eq_pf_lem_achievable_d1}, \eqref{eq_pf_lem_achievable_a1} and \eqref{eq_pf_lem_achievable_c2}, and incorporating the fact that $R_1 \geq 0$, we have respectively the two corresponding stricter bounds
\begin{IEEEeqnarray}{rl}
R_1\leq & \Big( \log \det \left(I_{N_1}+\rho_{11} H_{11}H_{11}^{\dagger}\right)-n_1^* \Big)^+ \quad {\rm and} \label{eq:intermediate-rate-region-a} \\
R_1\leq & \Big(\log \det \left(I_{M_1}+\rho_{11} H_{11}^\dagger H_{11}+K_1^{-1}\right)\nonumber\\
&~~~~-(m_{11}\log(M_1)+m_{12}\log(M_1+1))-\hat{m}_{21}\Big)^+~~~
\label{eq:intermediate-rate-region-b}
\end{IEEEeqnarray}
where $n_i^*$ is defined in \eqref{eq_def_nis}.

Similarly, the two bounds on $R_2$ from \eqref{eq:compact-gaussian-rate-region-1c} and \eqref{eq:compact-gaussian-rate-region-1d} imply (using the inequalities \eqref{eq_pf_lem_achievable_d2}, \eqref{eq_pf_lem_achievable_a2} and \eqref{eq_pf_lem_achievable_c1}) respectively that
\begin{IEEEeqnarray}{ll}
R_2 \leq & \Big( \log \det \left(I_{N_2}+\rho_{22} H_{22}H_{22}^{\dagger}\right)-n_2^*\Big)^+  \quad {\rm and} \label{eq:intermediate-rate-region-c}
\end{IEEEeqnarray}
\begin{IEEEeqnarray}{rl}
R_2 \leq & \Big(\log \det \left(I_{M_2}+\rho_{22} H_{22}^\dagger H_{22}+K_2^{-1}\right)\nonumber\\
&~~-(m_{22}\log(M_2)+m_{21}\log(M_2+1))-\hat{m}_{12}\Big)^+ .~~~~
\label{eq:intermediate-rate-region-d}
\end{IEEEeqnarray}

Recall from \eqref{eq_def_ni} that
\begin{equation*}
    n_i= \max \{(m_{ii}\log(M_i)+m_{ij}\log(M_i+1))+\hat{m}_{ji}, n_i^*\},
\end{equation*}
$\forall ~i\neq j\in \{1,2\}$, so that the two bounds on $R_1$ in \eqref{eq:intermediate-rate-region-a} and \eqref{eq:intermediate-rate-region-b} and the two bounds on $R_2$ in \eqref{eq:intermediate-rate-region-c} and \eqref{eq:intermediate-rate-region-d} can be combined into
\begin{IEEEeqnarray}{ll}
R_1\leq & \Big(\log \det \left(I_{N_1}+\rho_{11} H_{11}H_{11}^{\dagger}\right)-n_1\Big)^+ \quad {\rm and} \\
R_2\leq & \Big(\log \det \left(I_{N_2}+\rho_{22} H_{22}H_{22}^{\dagger}\right)-n_2\Big)^+
\end{IEEEeqnarray}
where we use the fact that $K_i^{-1}$ is a p.d. matrix and $\log\det(.)$ is a monotonically increasing function in the cone of p.s.d. matrices. The above bounds are the first two bounds in the set of bounds in \eqref{eq:rate-region-within-ni-bits1}-\eqref{eq:rate-region-within-ni-bits7} that were to be proved.

Consider next the sum rate bound from \eqref{eq:compact-gaussian-rate-region-1e}
\begin{IEEEeqnarray}{ll}
R_1 + R_2 \leq I(X_2^g,W_1^g;Y_2^g) + I(X_1^g;Y_1^g|W_1^g,W_2^g). \label{rewritecgrr1e}
\end{IEEEeqnarray}
Using \eqref{eq_pf_lem_achievable_g1} and \eqref{eq_pf_lem_achievable_a1}, it can be shown that
\begin{IEEEeqnarray*}{rl}
R_1+R_2\leq &\Big(\log \det \left(I_{N_2}+\rho_{12} H_{12}H_{12}^{\dagger}+\rho_{22} H_{22}H_{22}^{\dagger}\right)+ \nonumber\\
&\log \det \left(I_{N_1}+\rho_{11} H_{11}K_{1}H_{11}^{\dagger}\right)-(n_1^*+n_2^*)\Big)^+,
\end{IEEEeqnarray*}
which is the third bound in the set of bounds in \eqref{eq:rate-region-within-ni-bits1}-\eqref{eq:rate-region-within-ni-bits7} to be proved.
The last four bounds in \eqref{eq:rate-region-within-ni-bits1}-\eqref{eq:rate-region-within-ni-bits7} can be similarly proved starting from
\eqref{eq:compact-gaussian-rate-region-1e}-\eqref{eq:compact-gaussian-rate-region-1i} using some of the inequalities developed in \eqref{eq_pf_lem_achievable_a1}-\eqref{eq_pf_lem_achievable_e2}. The details are left to the reader.

Since the bounds that define the rate region $\mathcal{R}^{G_e}_{\textrm{HK}}(P_s)$ imply the bounds in \eqref{eq:rate-region-within-ni-bits1}-\eqref{eq:rate-region-within-ni-bits7} that define $\mathcal{R}_a(\mathcal{H},\bar{\rho})$, we have that $ \mathcal{R}_a(\mathcal{H},\bar{\rho}) \subseteq \mathcal{R}^{G_e}_{\textrm{HK}}(P_s)$ as was to be proved.

\section{Proof of Lemma~\ref{lem:equivalent-gaussian-rate-region}}
\label{app:gap-to-outerbound}
We prove that for any given $(R_1,R_2)\in \mathcal{C}\left(\mathcal{H},\bar{\rho}\right)$, the rate pair $((R_1-n_1)^+,(R_2-n_2)^+)\in \mathcal{R}_a\left(\mathcal{H},\bar{\rho}\right)$. We prove this by contradiction. Using Lemma \ref{lem_upper_bound}, $ (R_1,R_2)\in \mathcal{C}\left(\mathcal{H},\bar{\rho}\right) $ implies that $  (R_1,R_2)\in \mathcal{R}^u\left(\mathcal{H},\bar{\rho}\right) $. With $\hat{R}_i= (R_i-n_i)^+$ for $i=1,2$, suppose that $(\hat{R}_1,\hat{R}_2)\notin \mathcal{R}_a\left(\mathcal{H},\bar{\rho}\right)$. Without loss of generality, we assume that $R_i\geq n_i,~\forall~ i$ because the other case follows trivially. Since one or more of the bounds of Lemma~\ref{lem:rate-region-within-ni-bits} are violated by $(\hat{R}_1,\hat{R}_2)$, let us suppose that the third bound is violated, i.e.,
\begin{IEEEeqnarray*}{rl}
(\hat{R}_1+\hat{R}_2)=&\left(R_1+R_2-(n_1+n_2)\right), \\
>  & ~ \log \det \left(I_{N_2}+\rho_{12} H_{12}H_{12}^{\dagger}+\rho_{22} H_{22}H_{22}^{\dagger}\right)\\
+&\log \det \left(I_{N_1}+\rho_{11} H_{11}K_{1}H_{11}^{\dagger}\right)-(n_1^*+n_2^*), \\
\geq & ~ \log \det \left(I_{N_2}+\rho_{12} H_{12}H_{12}^{\dagger}+\rho_{22} H_{22}H_{22}^{\dagger}\right)\\
+&\log \det \left(I_{N_1}+\rho_{11} H_{11}K_{1}H_{11}^{\dagger}\right)-(n_1+n_2),
\end{IEEEeqnarray*}
where we used the fact that $n_i^* \leq n_i$.
This implies that
\begin{IEEEeqnarray*}{rl}
\left(R_1+R_2\right)>  & ~ \log \det \left(I_{N_2}+\rho_{12} H_{12}H_{12}^{\dagger}+\rho_{22} H_{22}H_{22}^{\dagger}\right)\\
 &+\log \det \left(I_{N_1}+\rho_{11} H_{11}K_1H_{11}^{\dagger}\right).
\end{IEEEeqnarray*}
which means that $(R_1,R_2)\notin \mathcal{R}^u\left(\mathcal{H},\bar{\rho}\right)$, clearly a contradiction. A contradiction similarly results more generally from $(\hat{R}_1,\hat{R}_2)$ violating one or more bounds of Lemma~\ref{lem:rate-region-within-ni-bits}.

\section{Proof of Lemma~\ref{lem_reciprocity_ub}}
\label{pf_lem_reciprocity_ub}
We shall prove this lemma in two steps. In step one, we shall prove
\begin{equation*}
\mathcal{R}^u\left(\mathcal{H}, \bar{\rho}\right)=\mathcal{R}^u\left(\tilde{\mathcal{H}}, \bar{\rho}^r\right),
\end{equation*}
where $\tilde{\mathcal{H}}=\{H_{11}^\dagger, H_{21}^\dagger, H_{12}^\dagger, H_{22}^\dagger\}$ and in the second step we shall prove that
\begin{equation*}
    \mathcal{R}^u\left(\tilde{\mathcal{H}}, \bar{\rho}^r\right)=\mathcal{R}^u\left(\mathcal{H}^r, \bar{\rho}^r\right).
\end{equation*}
Clearly, the above two equalities prove the lemma.

{\it Step1}: Denote the right hand sides of the seven upper bounds in \eqref{eq_bound1}-\eqref{eq_bound7} that constitute the outer bound $\mathcal{R}^u\left(\mathcal{H}, \bar{\rho}\right) $ as $I_{bk}$ for $k \in \{1,2, \cdots , 7\}$. Let us consider the interference channel $\mathcal{IC}\left(\tilde{\mathcal{H}}, \bar{\rho}^r\right)$. Following a similar method as in Lemma~\ref{lem_upper_bound} we can derive an upper bound to the capacity region of this IC. Let the corresponding bounds of $\mathcal{R}^u\left(\tilde{\mathcal{H}}, \bar{\rho}^r\right)$ be denoted by $I_k^r$, $1\leq k\leq 7$. In what follows, we shall first prove that $I_{b3}=I_4^r$, $I_{b4}=I_3^r$ and $I_{bk}=I_k^r$ for $k\in\{1,2,5,6,7\} $.

The equality $I_{b3}=I_4^r$ is proved next. From \eqref{eq_bound3}, we get equation \eqref{eq:Ib3eqIr4} as shown at the top of the next page,
\begin{figure*}
\begin{center}
\begin{IEEEeqnarray}{rl}
I_{b3}=&\log \det \Big[I_{N_2}+\rho_{12} H_{12}H_{12}^{\dagger}+\rho_{22} H_{22}H_{22}^{\dagger}\Big]+\log \det \Big[I_{N_1}+\rho_{11} H_{11}\Big[I_{M_1}+\rho_{12}H_{12}^\dagger H_{12}\Big]^{-1}H_{11}^{\dagger}\Big],\nonumber\\
=&\log \det \Big[I_{N_2}+\rho_{12} H_{12}H_{12}^{\dagger}+\rho_{22} H_{22}H_{22}^{\dagger}\Big]
+\log \det \Big[I_{M_1}+\rho_{12}H_{12}^\dagger H_{12}+\rho_{11} H_{11}^\dagger H_{11}\Big]\nonumber\\
& \qquad \qquad \qquad \qquad \qquad \qquad \qquad \qquad \qquad \qquad -\log \det \Big[I_{M_1}+\rho_{12}H_{12}^\dagger H_{12}\Big],\nonumber\\
\stackrel{(a)}{=}&\log \det \Big[I_{N_2}+\rho_{12} H_{12}H_{12}^{\dagger}+\rho_{22} H_{22}H_{22}^{\dagger}\Big]
+\log \det \Big[I_{M_1}+\rho_{12}H_{12}^\dagger H_{12}+\rho_{11} H_{11}^\dagger H_{11}\Big]\nonumber\\
& \qquad \qquad \qquad \qquad \qquad \qquad \qquad \qquad \qquad \qquad -\log \det \Big[I_{N_2}+\rho_{12}H_{12} H_{12}^\dagger\Big],\nonumber\\
=& \log \det \Big[I_{M_2}+\rho_{22}H_{22}^\dagger\Big[I_{N_2}+\rho_{12} H_{12}H_{12}^{\dagger}\Big]^{-1}H_{22}\Big]
+\log \det \Big[I_{M_1}+\rho_{12}H_{12}^\dagger H_{12}+\rho_{11} H_{11}^\dagger H_{11}\Big]=I_4^r.\label{eq:Ib3eqIr4}
\end{IEEEeqnarray}
\end{center}
\hrule
\end{figure*}
where step $(a)$ follows from the fact that $\log\det(I+AB)=\log\det(I+BA)$. Similarly, it can be proved that $I_{b4}=I_{3}^r$. The equality of the first two bounds, namely, $I_{b1}=I_1^r$ and $I_{b2}=I_2^r$ follow trivially from the identity $\log\det(I+AB)=\log\det(I+BA)$. Now, the first part of the fifth bound is proved in equation \eqref{eq:Ib51eqIr52} at the top of the next page.
\begin{figure*}
\begin{center}
\begin{IEEEeqnarray}{rl}
I_{b5}(1) \triangleq \log \det &\Big[I_{N_1}+ \rho_{21} H_{21}H_{21}^{\dagger}+ \rho_{11} H_{11}K_{1}H_{11}^{\dagger}\Big],\nonumber\\
=\log \det &\Big[I_{N_1}+  \rho_{11}\Big[I_{N_1}+ \rho_{21} H_{21}H_{21}^{\dagger}\Big]^{-1} H_{11}K_{1}H_{11}^{\dagger}\Big]+\log\det\Big[I_{N_1}+ \rho_{21} H_{21}H_{21}^{\dagger}\Big],\nonumber\\
=\log \det &\Big[I_{M_1}+  \rho_{11}H_{11}^{\dagger}\Big[I_{N_1}+ \rho_{21} H_{21}H_{21}^{\dagger}\Big]^{-1} H_{11}K_{1}\Big]+\log\det\Big[I_{M_2}+ \rho_{21} H_{21}^\dagger H_{21}\Big],\nonumber\\
=\log \det &\Big[K_1^{-1}+  \rho_{11}H_{11}^{\dagger}\Big[I_{N_1}+ \rho_{21} H_{21}H_{21}^{\dagger}\Big]^{-1} H_{11}\Big]+\log\det(K_1)+\log\det\left(K_2^{-1}\right),\nonumber\\
=\log \det &\Big[I_{M_1}+ \rho_{12}H_{12}^\dagger H_{12}+ \rho_{11}H_{11}^{\dagger}\Big[I_{N_1}+ \rho_{21} H_{21}H_{21}^{\dagger}\Big]^{-1} H_{11}\Big] +\log\det(K_1)-\log\det\left(K_2\right).\label{eq:Ib51eqIr52}
\end{IEEEeqnarray}
\end{center}
\hrule
\end{figure*}
Similarly, it can be easily proved that
\begin{IEEEeqnarray}{rl}
I_{b5}(2) \triangleq &\log \det \left(I_{N_2}+ \rho_{12} H_{12}H_{12}^{\dagger}+ \rho_{22} H_{22}K_{2}H_{22}^{\dagger}\right)\nonumber\\
=& \log \det \Big[I_{M_2}+ \rho_{21}H_{21}^\dagger H_{21}\nonumber\\
&~~~~~+\rho_{22}H_{22}^{\dagger}\left(I_{N_2}+ \rho_{12} H_{12}H_{12}^{\dagger}\right)^{-1} H_{22}\Big]\nonumber\\
&~~~~~ +\log\det(K_2)-\log\det(K_1).\label{eq:Ib52eqIr51}
\end{IEEEeqnarray}
Combining equations \eqref{eq:Ib51eqIr52} and \eqref{eq:Ib52eqIr51} we get
\begin{IEEEeqnarray*}{rl}
I_{b5} = &I_{b5}(1)+I_{b5}(2),\\
= &\log \det \left(I_{M_1}+ \rho_{12}H_{12}^\dagger H_{12}+ \rho_{11}H_{11}^{\dagger}(K_1^r)^{-1} H_{11}\right)\\
&+ \log \det \left(I_{M_2}+ \rho_{21}H_{21}^\dagger H_{21}+ \rho_{22}H_{22}^{\dagger}\left(K_2^r\right)^{-1} H_{22}\right),\\
=&I_5^r,
\end{IEEEeqnarray*}
where $K_i^r=\left(I_{N_i}+ \rho_{ji} H_{ji}H_{ji}^{\dagger}\right)$ for $1\leq i\neq j\leq 2$. Proving the equality of the other two bounds is similar. Hence, the set upper bounds for the capacity region of $\mathcal{IC}\left(\tilde{\mathcal{H}}, \bar{\rho}^r\right)$ defines the same set of rate pairs as $\mathcal{R}^u\left(\mathcal{H},\bar{\rho}\right)$.

{\it Step2}: Suppose $S$ is a p.s.d. matrix and $S^*$ represents its complex conjugate, i.e., the matrix obtained by replacing all its entries by the corresponding complex conjugates. Then, using the fact that its eigen-values are real, it can be easily be proved that $ \log\det(I+S)=\log\det(I+S^*) $. However, note that all the terms in the different bounds of Lemma~\ref{lem_upper_bound} are of the form just described. This in turn proves that if we replace all the channel matrices of a two-user MIMO IC by their complex conjugates the set of upper bounds remain the same. From this fact, it easily follows that $ \mathcal{R}^u\left(\tilde{\mathcal{H}}, \bar{\rho}^r\right)=\mathcal{R}^u\left( \mathcal{H}^r, \bar{\rho}^r\right) $.

\bibliographystyle{IEEEtran}
\bibliography{mybibliography}

\vspace{1cm}


\end{document}